\documentclass[11pt,showpacs,preprint,preprintnumbers,nofootinbib,
superscriptaddress,amsmath,amssymb]{revtex4}

\usepackage{graphicx}
\usepackage{dcolumn}
\usepackage{bm}
\usepackage{amssymb}
\usepackage{amsmath}
\usepackage{epsfig}    
\usepackage{color}
\usepackage{slashed}

\usepackage{float}

\newcommand{\PI}{\textrm{PI}}

    \def\slashchar#1{\setbox0=\hbox{$#1$} 
    \dimen0=\wd0 
    \setbox1=\hbox{/} \dimen1=\wd1 
    \ifdim\dimen0>\dimen1 
    \rlap{\hbox to \dimen0{\hfil/\hfil}} 
    #1 
    \else 
    \rlap{\hbox to \dimen1{\hfil$#1$\hfil}} 
    / 
 \fi}

\begin{document} 

\title{Testing the dark matter scenario in the inert doublet model by future precision measurements of the Higgs boson couplings}

\preprint{UT-HET 114}
\author{Shinya Kanemura}
\email{kanemu@sci.u-toyama.ac.jp}
\affiliation{Department of Physics, University of Toyama, \\3190 Gofuku, Toyama 930-8555, Japan}
\author{Mariko Kikuchi}
\email{kikuchi@jodo.sci.u-toyama.ac.jp}
\affiliation{Department of Physics, University of Toyama, \\3190 Gofuku, Toyama 930-8555, Japan}
\author{Kodai Sakurai}
\email{sakurai@jodo.sci.u-toyama.ac.jp}
\affiliation{Department of Physics, University of Toyama, \\3190 Gofuku, Toyama 930-8555, Japan}

\begin{abstract}
  We evaluate radiative corrections to the Higgs boson couplings in the inert doublet model,
  in which the lightest component of the $Z_2^{}$ odd scalar doublet field can be a dark matter candidate. 
  The one-loop contributions to the $hVV$, $hff$ and $hhh$ couplings are calculated in the on-shell scheme,
  where $h$ is the Higgs boson with the mass 125 GeV, $V$ represents a weak gauge boson and $f$ is a fermion. 
  We investigate how the one-loop corrected Higgs boson couplings can be deviated from the predictions in the standard model under the
  constraints from perturbative unitarity and vacuum stability in the scenario where the model can explain current dark matter data. 
  When the mass of the dark matter is slightly above a half of the Higgs boson mass, 
  it would be difficult to test the model by the direct search experiments for dark matter.   
  We find that in such a case 
  the model can be tested at future collider experiments by either the direct search of heavier inert particles 
  or precision measurements of the Higgs boson couplings.  
 \end{abstract}
 
 \pacs{12.60.-i, 14.80.Cp, 12.15.Lk, 95.35.+d}
 
\maketitle
\clearpage

\section{Introduction}

Inspite that the standard model (SM) has brought great success in describing the nature of particles,
there are still some phenomena which cannot be explained in this model, 
such as dark matter, neutrino oscillation and baryon asymmetry of the universe. 
At the same time we know nothing about the real shape of the Higgs sector, 
though a Higgs boson ($h$) was found in 2012~\cite{LHC_ATLAS, LHC_CMS} and its basic properties turned out to be similar to those of the Higgs boson in the SM~\cite{LHC_Higgs_mass, ATLAS_CMS_h_coup}. 
Therefore, one can consider the possibility that the above phenomena are explained by introducing an extended version of the Higgs sector.

The Weakly Interacting Massive Particle (WIMP) is 
a promising scenario for dark matter, in which the mass of the dark matter particle is near the electroweak scale. 
The inert doublet model (IDM)~\cite{IDM_VS1,LEP_IDM} is one of the simplest models for the WIMP dark matter scenario,
in which an iso-spin doublet scalar field is added to the SM Higgs sector and the field is assumed to be
odd under an unbroken discrete $Z_2^{}$ symmetry. 
After electroweak symmetry breaking, there are four $Z_2^{}$ odd scalar states; i.e., the CP-even $H$, the CP-odd $A$ and the charged $H^\pm$ 
scalar bosons, and 
the lightest component can be a dark matter candidate when it is electrically neutral. 

The dark matter scenario in the IDM has been tested by various experiments for dark matter searches
such as LUX~\cite{LUX}, SuperCDMS~\cite{s-CDMS}, Fermi-LAT~\cite{Fermi-LAT} and AMS-02~\cite{AMS-02}. 
By the direct search experiments, the mass of dark matter in the IDM has been currently constrained to be
around a half of the mass (125 GeV) of the Higgs boson or above about 500 GeV~\cite{IDM_DM_th3, Compressing_IDM, chile_DM}. 
In the future, the resonance region of the former case might not be completely excluded~\cite{abe_sato} by the direct search experiments~\cite{X1T, LZ}. 

Collider experiments can also be a useful tool to test the dark matter scenario in the IDM. 
The phenomenology of the IDM at hadron colliders has been studied in the literature~\cite{IDM_DM_th8, IDM_DM_th9, Compressing_IDM, LHC_direct_Cao, LHC_direct_Su, LHC_run-I_bound1, LHC_run-I_bound2},   
where final states such as a dilepton/dijet  signal 
with missing $E_T^{}$ following $H \to AZ^{(\ast)}$ ($H^\pm \to AW^\pm$) from $HA$ ($H^+H^-$) 
production via the Drell-Yan processes have mainly been analyzed, 
assuming that $A$ is the dark matter candidate. 
The mass determination of the dark matter has been discussed at future lepton colliders~\cite{IDM_ILC1, IDM_ILC2}. 

In this paper, we investigate how the dark matter scenario can be tested by the precision measurements of the couplings 
of the discovered Higs boson with the mass 125 GeV at future collider experiments. 
To this end, we evaluate radiative corrections to the $hVV$, $hff$ and $hhh$ couplings at the one-loop level in the on-shell scheme, 
where $V$ represents a weak gauge boson and $f$ is a fermion. 
The calculation has been done as a part of the H-COUP project~\cite{H-COUP}, where a full set of the fortran codes to compute the Higgs boson couplings 
at the one-loop level is being prepared in various extended Higgs sectors such as the Higgs singlet model~\cite{HSM_KKY}, the two Higgs doublet models (2HDMs)~\cite{2HDM_Yukawa_KKY, 2HDM_full}, the Higgs triplet model~\cite{HTM_reno, HTM_reno2} etc~\cite{MSSM_reno_h,HSM_reno_h}.
Although the one-loop corrections to the Higgs boson couplings in the IDM have been studied in Ref.~\cite{idm_hhh_ref}, 
our results partially do not agree with their results.  
We study deviations from the SM predictions in these Higgs boson couplings in the dark matter scenarios where the mass of dark matter is taken to be 63 GeV
and 500 GeV. 
We find that even in the case where the mass of the dark matter is slightly above a half of the Higgs boson mass, 
the model can be tested at future collider experiments by either the direct search of heavier inert particles 
or precision measurements of the Higgs boson couplings.  

This paper is organized as follows.
In Sec.~II the IDM is briefly reviewed, 
and its dark matter scenarios are discussed in Sec.~III.
Detailed explanation for one-loop calculations of the Higgs boson couplings is given in Sec.~IV.
Numerical evaluations for the scaling factors of the Higgs boson couplings are presented in the dark matter scenarios in Sec.~V.
Conclusions are given in Sec.~VI.
Exact formulae for the one-loop corrected Higgs boson couplings are listed in Appendix.

\section{Model}

The scalar sector of the IDM contains a complex isospin doublet field $\Phi_2^{}$ with hypercharge $Y=1/2$ in addition to the SM Higgs doublet field $\Phi_1^{}$.
The inert doublet field $\Phi_2^{}$ is odd under a discrete $Z_2^{}$ symmetry.
Under the electroweak symmetry $SU(2)\times U(1)$ and the unbroken discrete $Z_2^{}$ symmetry, the Higgs potential is given by
\begin{align}
	V =&\mu^2_1|\Phi_1|^2+\mu^2_2|\Phi_2|^2
			+\frac{1}{2}\lambda_1|\Phi_1|^4+\frac{1}{2}\lambda_2|\Phi_2|^4  \nonumber\\ 
			&+\lambda_3|\Phi_1|^2|\Phi_2|^2
			+\lambda_4|\Phi_1^\dagger\Phi_2|^2
			+\frac{1}{2}\{ \lambda_5(\Phi_1^\dagger\Phi_2)^2+h.c.\},   \label{eq:potential}
\end{align}
where all parameters can be takes to be real.
The coupling constants 
are bounded by several theoretical constraints such as vacuum stability~\cite{IDM_VS1, IDM_VS2} and perturbative unitarity~\cite{Kanemura:1993hm, Akeroyd:2000wc}. 
The requirement of vacuum stability is as provided below,
\begin{align}
  \lambda_1^{} > 0,\ \lambda_2^{} > 0,\ \sqrt{\lambda_1^{} + \lambda_2^{}} + \lambda_3^{} > 0,\ \lambda_3^{}+\lambda_4^{} \pm |\lambda_5^{}| > 0. \label{eq:vs}
\end{align}
We evaluate constraints on the coupling constants from perturbative unitarity adopting the formulae given in Refs.~\cite{Kanemura:1993hm,Akeroyd:2000wc}.

The doublet scalar fields are parameterized as 
\begin{equation}
	\Phi_1^{}=
	\begin{pmatrix}
 		G^+\\
 		\frac{1}{\sqrt{2}}(h+v+i G^0)
	\end{pmatrix}
	,\  \Phi_2^{}=
	\begin{pmatrix}
 		H^+\\
 		\frac{1}{\sqrt{2}}(H+iA)
	\end{pmatrix},
\end{equation}
where $h$ is the SM Higgs boson with the mass 125 GeV, $v$ is the vacuum expectation value (VEV), 
and $G^+$ and $G^0$ are Nambu-Goldstone bosons which are absorbed into the longitudinal components of the W and Z bosons. 
We call additional ($Z_2^{}$ odd) scalar bosons $H, A, H^{\pm}$ the inert scalar bosons. 

After imposing the stationary condition, masses of scalar bosons are given by   
\begin{align}
	m_h^2&=\lambda_1 v^2, \label{eq:mhsq}\\
	m^2_{H^+}&=\mu_2^2+\frac{1}{2}\lambda_3 v^2, \label{eq:mHpsq}\\
	m_H^2&=\mu_2^2+\frac{1}{2}(\lambda_3+\lambda_4+\lambda_5)v^2, \label{eq:mHsq}\\
	m_A^2&=\mu_2^2+\frac{1}{2}(\lambda_3+\lambda_4-\lambda_5)v^2. \label{eq:mAsq}	
\end{align}
In this paper, we choose $v, m_h, \mu_2^2, \lambda_2, m_A, m_H$ and $m_{H^{\pm}}$ as input parameters. 
Coupling constants for the vertices $h\phi\phi$ can be given by
 \begin{align}
	\lambda_{hHH}=-\frac{m_H^2-\mu^2_2}{\upsilon}, \ 
	\lambda_{hAA}=-\frac{m_A^2-\mu^2_2}{\upsilon}, \ 
	\lambda_{hH^{+}H^{-}}=-2\frac{m^2_{H^+}-\mu^2_2}{\upsilon}, \\
	\lambda_{hhh}=-\frac{m^2_h}{2\upsilon} , \ 
	\lambda_{hG^0G^0}=-\frac{m^2_h}{2\upsilon} , \ 
	\lambda_{hG^+G^-}=-\frac{m^2_h}{\upsilon}, \ \ \ \ \ \ \ \ \ \ \ \ \ \ 
 \end{align}
 where the definision of $\lambda_{h\phi\phi}^{}$ is as follows, 
 \begin{align}
   \mathcal{L} = \lambda_{h\phi\phi} h\phi\phi + \cdots. 
    \end{align}

 We take into account experimental constraints given by several collider 
 experiments such as precision measurements of electroweak oblique corrections and 
direct searches of the inert scalar bosons. 
The data of precision measurements of the electroweak oblique corrections at LEP~\cite{LEP_Z} give bounds on masses of  
the inert scalar bosons, because loop corrections of an additional scalar doublet field 
contributes to the electroweak $S$, $T$ and $U$ parameters~\cite{STU_th1, LEP_IDM, STU_th2}. 
A deviation from the SM prediction in the 
$T$ parameter $\Delta T$ is experimentally given by $\Delta T \simeq 0.07\pm 0.08$~\cite{LEP_IDM}. 
In the case with $m_H^{}\simeq m_A^{} \simeq m_{H^\pm}^{}$, $\Delta T$ can be approximately expressed by  
\begin{align}
 \Delta T \simeq \frac{1}{24\pi^2\alpha_\textrm{EM} v^2}(m_{H^{\pm}}-m_A)(m_{H^{\pm}}-m_H), 
\end{align}
where $\alpha_\textrm{EM}^{}$ is the fine structure constant of electromagnetic interaction. 
In addition, as it is useful for our later discussions, 
we give another approximate formula for $\Delta T$, 
which can be applied to the case with $m_A^{}\ll m_{H^\pm}^{}, m_H^{}$ 
as\footnote{Notice that the formula for $m_H^{}\ll m_{H^\pm}^{}, m_A^{}$ is obtained by 
the replacement of $A \leftrightarrow H$ in Eq.~(\ref{eq:delta_T}). } 
\begin{equation}
\Delta T \simeq \frac{1}{16\pi^2\alpha_\textrm{EM}^{} v^2}\left\{\frac{1}{2}(m_{H^{\pm}}^2- m_H^2) -\frac{2}{3}(m_{H^\pm}^{} - m_H)^2 \right\}.
\label{eq:delta_T}
 \end{equation}
In the both cases, the data of $\Delta T$ imply that a mass difference between $H^\pm$ and a neutral inert scalar boson ($\Delta m$) must not be too large, i.e. $\Delta m \lesssim\mathcal{O}(10)$ GeV.  
 Following regions have also been excluded by direct searches of inert scalar bosons at 
 LEP II~\cite{LEP-II-H, LEP-II-Hp}, 
 \begin{align}
 & m_{H^\pm} \lesssim 70\; {\rm to }\; 90 \textrm{GeV}, \\
 & m_A<80{\rm GeV},\ m_H<100{\rm GeV},\ m_H - m_A > 8{\rm GeV},
 \end{align}
 where we assume $m_A^{} < m_H^{}$.
 \color{black}

 \section{Dark Matter}
 
 The lightest neutral inert scalar boson (either $H$ or $A$) can be a candidate of dark matter. 
 In this paper, we assume that $A$ is the dark matter; i.e. $m_A^{} < m_H^{}$, which corresponds that $\lambda_5^{}$ is positive. 

 The data for the thermal relic abundance~\cite{WMAP-relic,Planck-relic} bound the mass region of dark matter to be 
 $3 \textrm{GeV} \lesssim 
 m_A^{} \lesssim 100$ GeV and $m_A^{} \gtrsim 500$ GeV under the assumption that only the lightest inert scalar boson is dark matter~\cite{IDM_DM_th1,IDM_DM_th2,IDM_DM_th4,IDM_DM_th5,IDM_DM_th6,IDM_DM_th7,IDM_DM_th8,IDM_DM_th9,Compressing_IDM,chile_DM}.  
 Moreover, the mass region has been narrowed down to $50 \textrm{GeV} \lesssim m_A^{} \lesssim 80$ GeV and $m_A^{} \gtrsim 500$ GeV 
 by the current data of direct detection of dark matter~\cite{LEP_IDM,IDM_DM_th1,IDM_DM_th2,IDM_DM_th3,IDM_DM_th4,IDM_DM_th5,IDM_DM_th6,IDM_DM_th7,IDM_DM_th8,IDM_DM_th9,Compressing_IDM,chile_DM}. 
 It is hard to completely explore the remaining region at future experiments of direct searches such as Xenon 1T~\cite{X1T} and LZ~\cite{LZ} and so on. 
 If there are other dark matter candidates in addition to the inert scalar boson,
 the region for $m_h^{}/2 \lesssim m_A^{} \lesssim 500$ GeV also can not be excluded by data of relic density~\cite{IDM_DM_th3,Compressing_IDM,chile_DM}.

In this paper, we are especially interested in the question whether or not 
it is possible to test the dark matter scenario in the challenging regions of direct searches 
by using future precision measurements of the Higgs boson couplings. 
In particular, we consider the following bench mark scenarios, 
 \begin{align}
   &\textrm{Scenario-A}\quad  m_A^{}=63 \textrm{GeV}, \mu_2^{2} = ( 61.50 \textrm{GeV})^2, m_H^{} = m_{H^\pm}^{}, \label{SA}\\
   &\textrm{Scenario-B}\quad  m_A^{}=500 \textrm{GeV}, \mu_2^{2} = ( 499.9 \textrm{GeV})^2, m_H^{} = m_{H^\pm}^{},  \label{SB2}
 \end{align}
where $m_H^{} = m_{H^\pm}^{}$ is assumed to satisfy the data of the T parameter~\cite{LEP_Z}. 
In TABLE~\ref{tab}, parameters in these scenarios are listed. 
Once we set $m_A^{}$ and $\mu_2^{2}$ to be those of Scenario-A, the value of $\lambda_A^{}$ is determined as, 
 \begin{align}
   &\textrm{Scenario-A}\quad    \lambda_{A}^{} \simeq 6.17 \times 10^{-3}, \\
    &\textrm{Scenario-B}\quad    \lambda_{A}^{} \simeq 4.97 \times 10^{-3}, 
   \end{align}
 due to Eq.~(\ref{eq:mAsq}),  where we define $\lambda_A^{} \equiv \lambda_3^{} +\lambda_4^{} - \lambda_5^{}$. 
In Scenario-B, the relic abundance of dark matter cannot be realized when the mass difference between $A$ and 
 $H^\pm$ is not small. In this case, we should suppose that there is another additional source of dark matter to satisfy the relic abundance.  
 
\begin{table}
   \caption{Parameter sets of bench mark scenarios}\label{tab}
   \begin{tabular}{|c||c|c|c||c|} 
     \hline
     & $m_A^{}$ [GeV] & $\mu_2^2$ [GeV$^2$] & $m_H^{}, m_{H^\pm}^{}$ & $\lambda_A^{}$ \\
     \hline\hline 
     Scenario-A & 63 & $(61.50)^2$ & $m_H^{} = m_{H^\pm}^{}$ &  $6.17 \times 10^{-3}$ \\
     \hline
     Scenario-B & 500 & $(499.9)^2$ & $m_H^{} = m_{H^\pm}^{}$ &  $4.97 \times 10^{-3}$ \\
     \hline
   \end{tabular}
   \end{table}

 In addition to Scenario-A and Scenario-B, 
 we define the third scenario just as a counter example, Scenario-C, 
 in which all inert scalar bosons are degenerated.
This scenario is not related to dark matter.

 \section{Calculation of one-loop corrections to the Higgs boson couplings}
 
In the IDM, 
the structure of counter-terms for the Higgs boson couplings is essentially 
the same as those in the SM~\cite{Hollik} , since there is no mixing between the SM fields and the additional inert fields. 
We here give the brief description for our renormalization scheme for completeness. 
 
 Regarding the Higgs potential in Eq.~(\ref{eq:potential}), 
 there are eight bare parameters;  i.e., $m_h^{}$, $v$, $T_h^{}$,   $m_{H}^{}$, $m_{A}^{}$, $m_{H^\pm}$, $\mu_2^{}$ and $\lambda_2$,  
 where $T_h^{}$ is the tatpole for $h$.  
 They are shifted to the renormalized parameters and the counter terms as 
 \begin{align}
   m_{h}^2     &\to m_{h}^2+\delta m_{h}^2, \\
   v                &\to v + \delta v, \\
   T_{h}          &\to \delta T_{h}, \\
   m_{\Phi}^2 &\to m_{\Phi}^2+\delta m_{\Phi}^2,\\
   \mu_2^2    &\to \mu_2^2+\delta \mu_2^2, \\ 
   \lambda_2 &\to\lambda_{2}+\delta \lambda_2. 
 \end{align}
 For our current task to calculate one-loop contributions to the Higgs boson couplings, 
 only three of them are used; i.e., $\delta m_h^2$, $\delta v$ and $\delta T_h$.  
 In addition, the wave function renormalization parameters is introduced for scalar fields as
 \begin{align}
   & \varphi  \to \left(1 + \frac{1}{2}\delta Z_\varphi^{} \right)\varphi, 
 \end{align}
 where $\varphi$ represents all scalar bosons, $\varphi = h, H, A, H^\pm, G^0$ and $G^\pm$. 
 
 The renormalization of the vacuum expectation value $\delta v$ is evaluated 
 by using the relation  
 \begin{align}
 \frac{\delta v}{v} = \frac{1}{2 m_W^2} \Pi_{WW,T}^{\rm 1PI}(0) + ({\rm vertex\; and\; box\; diagrams}), 
 \end{align}
 where the 1PI diagram contributions  $\Pi_{WW,T}^{\rm 1PI}(p^2)$ 
 to the W boson two point function in the IDM are given in Eqs. (\ref{2Ff_WW}) and (\ref{2FB_WW}) in Appendix. 
 In our calculation, we neglect the box diagram contribution 
 as they do not contain the one-loop contribution of inert scalars. 
 
 Renormalized tadpole parameter and two point functions for the Higgs boson $h$  are expressed by 
 \begin{align}
   \hat{T}_h&=\delta T_h+\Gamma_h^{\rm 1PI}, 
   \\\nonumber
   \hat{\Pi}_{hh}(p^2)&=\Pi_{hh}^{\rm 1PI}(p^2)
   +\left[(p^2-m_h^2)\delta Z_h-\delta m_h^2 \right]+\frac{\hat{T}_h}{\upsilon}, 
 \end{align}
 where $\Gamma_h^{1\PI}$ and $\Pi_{hh}^{1\PI}[p^2]$ are the one-loop contributions of 
 1PI diagrams.   
 We determine $\delta T_h^{}, \delta m_h^{2}$ and $\delta Z_h^{}$ by imposing the following renormalization conditions, 
 \begin{align}
   \hat{T}_h&=0, \quad
   \hat{\Pi}_{hh}(m^2_{h})=0,\quad  \left.\frac{d}{dp^2}\hat{\Pi}_{hh}(p^2)\right|_{p^2=m^2_{h}}=0. 
 \end{align} 
 Obtained counter terms are 
 \begin{align}
   \delta T_h&=-\Gamma^{\rm 1PI}_h,\quad
   \delta m_{h}^2=\Pi^{\rm 1PI}_{hh}(m_{h}^2)
   ,\quad 
   \delta Z_{h}=-\left.\frac{d}{dp^2}\Pi^{\rm 1PI}_{hh}(p^2)\right|_{p^2=m_{h}^2}, 
 \end{align}
 where the concrete expressions of $\Gamma^{\rm 1PI}_h$ and $\Pi^{\rm 1PI}_{hh}(p^2)$ 
 are given in Eqs. (\ref{1Ff_h}), (\ref{1FB_h}), (\ref{2Ff_hh}) and (\ref{2FB_hh}) in Appendix.

  The triple Higgs boson coupling is given at the tree level by  
 \begin{align}
 \Gamma_{hhh}^{\rm tree}=-\frac{3m_h^2}{\upsilon}, 
  \end{align} 
 and the one-loop corrected $hhh$ vertex is calculated as  
\begin{align}
  \label{hhh_sm}
   \hat{\Gamma}_{hhh}(p^2_1,p^2_2,q^2) =\Gamma_{hhh}^{\rm tree}+\delta \Gamma_{hhh} +\Gamma_{hhh}^{\rm 1PI}(p^2_1,p^2_2,q^2), 
 \end{align}  
where the counter term $\delta \Gamma_{hhh}$ is given by 
 \begin{align}
     \label{hhh_ana}
   &\delta{\Gamma}_{hhh} = -\frac{3m^2_h}{v}\left(\frac{\delta m^2_h}{m_h^2} -\frac{\delta v}{v}+\frac{3}{2}\delta Z_h \right),
 \end{align}
 and the 1PI diagram contribution $\Gamma_{hhh}^{\rm 1PI}$ 
 in the IDM is given in Eqs. (\ref{3Ff_hhh}) and (\ref{3FB_hhh}) in Appendix. 
  
 The $hVV$ ($V$ is a vector boson $W$ or $Z$) vertices are expressed in terms of the form factors as 
\begin{align}
\Gamma^{\mu\nu}_{hVV}&=\Gamma^{1}_{hVV}g^{\mu\nu}+\Gamma^{2}_{hVV}\frac{p_1^{\mu}p_2^{\nu}}{m_V^2}+i\Gamma^{3}_{hVV}\epsilon^{\mu\nu\rho\sigma}\frac{p_{1\rho}p_{2\sigma}}{m_V^2}, 
\end{align}
and the vertex function for the Higgs boson coupling with a fermion can be decomposed as 
\begin{align}
{\Gamma}^{}_{hff}&=\Gamma^{S}_{hff}+\gamma_5\Gamma^{P}_{hff}+\slashchar{p}_1\Gamma^{V_1}_{hff}
+\slashchar{p}_2\Gamma^{V_2}_{hff}
\\ 
 &+\slashchar{p}_1\gamma_5\Gamma^{A_1}_{hff}++\slashchar{p}_2\gamma_5\Gamma^{A_2}_{hff}+\slashchar{p}_1\slashchar{p}_2\Gamma^{T}_{hff}
+\slashchar{p}_1\slashchar{p}_2\gamma_5\Gamma^{PT}_{hff}.
\end{align}
They are given at the tree level by 
\begin{align}
&\Gamma_{hVV}^{\rm 1, {\rm tree}}=\frac{2 m_V^2}{\upsilon},  
\hspace{6mm}\Gamma_{hVV}^{\rm 2, {\rm tree}}=\Gamma_{hVV}^{\rm 3, {\rm tree}}=0, \\
&\Gamma_{hff}^{S,\rm tree}=-\frac{m_f}{\upsilon},  
\hspace{6mm}\Gamma_{hff}^{x, {\rm tree}}=0\ \ \ (x \neq S).
\end{align}
 At the one-loop level, each renormalized form factor can be expressed by
 \begin{align}
   \label{hVV_sm}
   &\hat{\Gamma}^{i}_{hVV}(p^2_1,p^2_2,q^2) = \Gamma_{hVV}^{i,{\rm tree}}+\delta \Gamma_{hVV}^{i}+\Gamma_{hVV}^{i,{\rm 1PI}}(p^2_1,p^2_2,q^2),
   \quad (i=1-3)
   \\
   \label{hff_sm}
   &\hat{\Gamma}^{x}_{hff}(p^2_1,p^2_2,q^2)=\Gamma_{hff}^{x, {\rm tree}}+\delta \Gamma_{hff}^{x}+\Gamma_{hff}^{x,{\rm 1PI}}(p^2_1,p^2_2,q^2),
   \quad (x=S, P, V_1, V_2, A_1, A_2, T, PT)
 \end{align}
 where $p_1^{}$ and $p_2^{}$ ($q$) in the form factors of $hX\overline{X}$ 
 are the incoming momenta of the particles $X$  and $\overline{X}$ (the outgoing momentum of the $h$ field).
 $\Gamma_{hXX}^\textrm{tree}$, $\delta \Gamma_{hXX}$ and $\Gamma_{hXX}^{1\PI}$ 
 are  the tree level contribution, the counter term and the one-loop 1PI diagram contributions to the $hX\overline{X}$ coupling, respectively.  
Explicit formulae for the 1PI diagram contributions 
$\Gamma^{1,1\PI}_{hVV}$ and $\Gamma^{S,1\PI}_{hff}$ are given 
 in Eqs.~(\ref{3Ff_hZZ}), (\ref{3FB_hZZ}), (\ref{3Ff_hWW}), (\ref{3FB_hWW}) and (\ref{3F_hff_S}) in Appendix.  
 Each $\delta \Gamma_{hXX}^{}$ is given in terms of the counter terms as  
 \begin{align}
   \label{hVV_1_ana}
   &\delta{\Gamma}^1_{hVV} = 
   +\frac{2m^2_V}{v}\left( \frac{\delta m_V^2}{m_V^2} -\frac{\delta v}{v}+	\frac{1}{2}\delta Z_h+\delta Z_V \right), \\
   &\delta\Gamma_{hVV}^2 =\delta\Gamma_{hVV}^3 = 0, \\
   \label{hff_S_ana}
   &\delta{\Gamma}^S_{hff}=-\frac{m_f}{\upsilon}\left( \frac{\delta m_f}{m_f}-\frac{\delta\upsilon}{\upsilon}
   +\frac{1}{2}\delta Z_h+\delta Z^f_V\right), \\
   &\delta\Gamma_{hff}^P = \delta\Gamma_{hff}^{V_1} = \delta\Gamma_{hff}^{V_2} = \delta\Gamma_{hff}^{A_1} =\delta\Gamma_{hff}^{A_2} =\delta\Gamma_{hff}^{T} =
   \delta\Gamma_{hff}^{PT} =0, 
 \end{align}
 where $\delta m_V^2$, $\delta Z_V$, $\delta m_f$ and $\delta Z_V^f$ 
 are the mass counter terms and the wave function renormalization factors 
 of weak gauge bosons and fermions, respectively, and they are given by 　　
 \begin{align}
\label{counterterm_m_w_m_z}
\delta m_{V}^2&={\rm }\Pi_{VV, T}^{{\rm 1PI}}(m_{V}^2),\ \ 
\delta Z_V=-\left.\frac{d}{dp^2}\Pi_{VV, T}^{{\rm 1PI}}(p^2)\right|_{p^2=m_V^2}, 
\end{align}
and 
 \begin{align}
	\frac{\delta m_f}{m_f}&=\Pi^{\rm 1PI}_{ff,V}(m_f^2)+\Pi^{\rm 1PI}_{ff,S}(m_f^2), \\
	\delta Z_V^f &= -\Pi^{\rm 1PI}_{ff,V}(m_f^2)-2m^2_f\Bigg[  
	\left.\frac{d }{d p^2}\Pi^{\rm 1PI}_{ff,V}(p^2)\right|_{p^2=m^2_f}
	+\left.\frac{d }{d p^2}\Pi^{\rm 1PI}_{ff,S}(p^2)\right|_{p^2=m^2_f}
	\Bigg],  
\end{align} 
where the concrete expressions of the 1PI diagram contributions to the gauge boson two point function 
$\Pi_{VV, T}^{{\rm 1PI}}(p^2)$ ($V=W$ and $Z$) and those to the fermion two point 
functions $\Pi^{\rm 1PI}_{ff,V}(p^2)$ and $\Pi^{\rm 1PI}_{ff,S}(p^2)$ are presented in Eqs.~(\ref{2Ff_WW}), (\ref{2FB_WW}), (\ref{2Ff_ZZ}), (\ref{2FB_ZZ}), (\ref{2F_ff_V}) and (\ref{2F_ff_S}) in Appendix, respectively. 
  
\section{Deviations on the Higgs couplings from the SM predictions at the one-loop level}

\subsection{The scaling factors}

The deviations in the Higgs boson couplings from the SM values are measured by introducing 
the scaling factors $\kappa_X^{}$ ($= 1 + \Delta \kappa_X^{}$). We define the one-loop corrected scaling factors in the IDM by 
 \begin{align}
   \label{kappa_ana_V}
   \Delta \hat{\kappa}_V(q^2)&=\frac{\hat{\Gamma}_{hVV}^1(m_V^2,m_h^2,q^2)_{{\rm IDM}}}{\hat{\Gamma}^{hVV}_1(m_V^2,m_h^2,q^2)_{{\rm SM} }}-1, \\
   \label{kappa_ana_f}
   \Delta \hat{\kappa}_f(q^2)&=\frac{\hat{\Gamma}^S_{hff}(m_f^2,m_f^2,q^2)_{{\rm IDM}}}{\hat{\Gamma}^S_{hff}(m_f^2,m_f^2,q^2)_{{\rm SM}}}-1, \\
   \label{kappa_ana_h}
   \Delta \hat{\kappa}_h(q^2)&=\frac{\hat{\Gamma}_{hhh}(m_h^2,m_h^2,q^2)_{{\rm IDM}}}{\hat{\Gamma}_{hhh}(m_h^2,m_h^2,q^2)_{{\rm SM}}}-1. 
 \end{align}
 In the following discussions, we set the value of $q^2$ to be $(m_V^{} + m_h^{})^2$, $m_h^2$ and $(2m_h^{})^2$ for the $hVV$, $hf\bar{f}$ and $hhh$ couplings, respectively. 
 We also numerically calculate the deviation on the decay rate of the process $h\to \gamma\gamma$ from the SM prediction 
 at the one-loop level~\cite{IDM_DM_th7,Arhirb_hgg,Maria_Bogumula_hgg}. 
 The deviations on the effective coupling $h\gamma\gamma$ is defined as 
 \begin{align}
   \Delta \hat{\kappa}_\gamma^{} \equiv \sqrt{\frac{\Gamma[h\to\gamma\gamma]_{\rm IDM}^{}}{\Gamma[h\to \gamma\gamma]_{\rm SM}^{}}} -1. 
 \end{align}

We here give several approximate formulae for the one-loop corrected Higgs boson couplings.
First of all, we consider the case where all inert scalar bosons are degenerated in mass, namely 
the case corresponding to Scenario-C. 
 If we expand $\Delta\hat{\kappa}_X^{}$ by $\epsilon$ ($\equiv m_h^{2}/m_{H^\pm}^{2} \ll 1$),
 we obtain the following approximate analytic formulae;
\begin{align}
  & 16\pi^2\Delta \hat{\kappa}_Z^{} \simeq 16 \pi^2 \Delta \hat{\kappa}_f^{}\simeq -\sum_{\Phi=A,H,H^\pm} c_\Phi^{}\frac{1}{6}\frac{m_{\Phi}^2}{v^2}
  \left(1-\frac{\mu_2^2}{m_\Phi^2}\right)^2+ \mathcal{O}(\epsilon),  \label{eq:dkz_appro1}\\
  & 16\pi^2\Delta \hat{\kappa}_h^{} \simeq \sum_{\Phi=A, H,H^\pm} c_\Phi^{}\frac{4}{3}
  \frac{m_\Phi^4}{m_h^2 v^2}\left(1 - \frac{\mu_2^2}{m_\Phi^2}\right)^3+ \mathcal{O}(\epsilon) \label{eq:dkh_appro1},
\end{align}
where $c_\Phi^{} = 2$ ($1$) for $\Phi=H^\pm$ ($H, A$). 
The masses of inert scalar bosons are given in Eqs.~(\ref{eq:mHpsq}) to (\ref{eq:mAsq}), which 
take a common shape as 
\begin{align}
   m_\Phi^2 = \mu_2^2 + {\cal O(}\lambda_i) v^2. 
\end{align}
If $m_\Phi$ is large because of a large $\mu_2^2$ ($\gg v^2$), then 
$m_\Phi^2 \simeq \mu_2^2$ so that the one-loop contributions given 
in Eqs.~(\ref{eq:dkz_appro1}) and (\ref{eq:dkh_appro1}) are suppressed  
by $1/m_\Phi^2$ in the large mass limit. 
On the other hand,  when $\mu_2^2 \sim v^2$, the one-loop contributions in 
Eqs.~(\ref{eq:dkz_appro1}) and (\ref{eq:dkh_appro1}) 
take positive power-like contributions; i. e., 
in proportion to $m_\Phi^2$ for $\Delta \hat{\kappa}_Z$ and $\Delta \hat{\kappa}_f$ 
and to $m_\Phi^4$ for $\Delta \hat{\kappa}_h$.  
In such a case, the scaling factors can be significantly large due to 
these non-decoupling contributions. 
In particular,  $\Delta \hat{\kappa}_h$ can easily be larger than 100\%, as 
previously pointed out in the context of the ordinary 2HDM or other extended Higgs sectors 
in Refs.~\cite{2HDM_full, HSM_KKY}. 

Next, we give another approximate formulae with $m_H^{} \simeq m_{H^\pm}^{}$ 
corresponding to Scenario-A and Scenario-B as follows, 
\begin{align}
  16\pi^2 \Delta\hat{\kappa}_Z^{} &\simeq 16\pi^2 \Delta\hat{\kappa}_f
  -\frac{1}{v^2}\left\{ \frac{2}{3}(m_H^{}-m_A^{})^2 -\frac{1}{30m_H^2}(m_H^{}-m_A^{})^4 \right\} \notag\\
  & + \frac{m_H^2}{v^2}\left(1-\frac{\mu_2^2}{m_H^2}\right)\left\{ \frac{2}{3}\left(1-\frac{m_A^{}}{m_H^{}}\right) -\frac{1}{15}\left(1-\frac{m_A^{}}{m_H^{}}\right)^3 \right\}
  +\mathcal{O}(\epsilon') , \label{eq:dkz_appro2}  \\
    16\pi^2 \Delta\hat{\kappa}_f^{} &\simeq -\sum_{\Phi=A,H,H^\pm} c_\Phi^{}\frac{1}{6}\frac{m_{\Phi}^2}{v^2}
  \left(1-\frac{\mu_2^2}{m_\Phi^2}\right)^2 + \frac{1}{v^2} \left\{ 
  \frac{1}{3} \left(m_{H^\pm}^{} - m_A\right)^2 -\frac{1}{60} \frac{\left(m_{H^\pm}-m_A\right)^4}{m_{H^\pm}^2}
  \right\} 
  +\mathcal{O}(\epsilon'),  \label{eq:dkb_appro2}\\
  16\pi^2 \Delta \hat{\kappa}_h^{} &\simeq \sum_{\Phi=H,H^\pm} c_\Phi^{}\frac{4}{3}
  \frac{m_\Phi^4}{m_h^2 v^2}\left(1 - \frac{\mu_2^2}{m_\Phi^2}\right)^3 +\mathcal{O}(\epsilon') ,
\end{align}
where $\epsilon' = m_h^{2}/m_{H^\pm}^{2}$. 
These approximate formulae are derived by expanding with the limit as $\epsilon' \ll 1$.
For Scenario-A, we see that loop corrections due to the inert particles other than $A$ always 
appear taking the form proportional to $m_{\Phi}^2$ for $\Delta\hat{\kappa}_Z^{}$ 
and  $\Delta\hat{\kappa}_f^{}$ and to $m_{\Phi}^4$ for $\Delta\hat{\kappa}_h$, because    
 $\mu_2^2$ is set to be about (61.5 GeV)$^2$ ($\ll v^2$) and $m_A$ is 63 GeV.   
Therefore, due to the non-decoupling effects, 
we have relatively large deviations in the Higgs boson coupling 
from the SM values in Scenario-A for relatively large values of $m_\Phi$.   
In Scenario-B, we also can have similar large deviations when $m_\Phi \gg 500$ GeV as 
we show numerically later.

In the limit $m_{H^\pm}^{2}\gg m_h^2$, 
$\Gamma[h\to\gamma\gamma]$ can be approximately expressed as,
\begin{align}
  \Gamma [h\to\gamma \gamma] \simeq \frac{\sqrt{2}G_F\alpha_{\rm EM}^2m_h^3}{64\pi^3}\left|-\frac{1}{6}\Big(1-\frac{\mu_2^2}{m^2_{H^{\pm}}} \Big)+\sum_fQ_f^2N_c^fI_f[m_h^2]+I_W[m_h^2] \right|^2,
  \label{eq:appro_dkg}
\end{align}
where the first, the second and the third terms are cntributions from the $H^\pm$ loop, fermion loops and the W boson loop, respectively. 
The loop functions $I_f[p^2]$ and $I_W[p^2]$ are given in Eqs. (\ref{hgamgam_f}) and (\ref{hgamgam_V})  in Appendix, respectively. 
The W boson loop contiribution is dominant 
in $\Gamma[h\to \gamma\gamma]$ in the SM, 
and $I_W^{}[m_h^{}]$ is a positive value. 
When $\mu_2^2$ is smaller than $m_{H^\pm}^2$, the $H^\pm$ loop contribution becomes destructive to the value of $\Gamma[h\to \gamma\gamma]$ of the SM.
We can see that if $\mu_2^2/m_{H^\pm}^2$ approaches to unity, 
the loop contribution of $H^\pm$ vanishes due to the decoupling property. 
If the case with $\mu_2^2 > m_{H^\pm}^{2}$ is realized, though rather unnatural, the prediction 
for $\Gamma[h\to \gamma\gamma]$ becomes larger than that of the SM; i.e.
$\Delta\kappa_\gamma^{} > 0$~\cite{Arhirb_hgg}.

\subsection{Numerical Evaluation}

We here show our numerical results of $\Delta \hat{\kappa}_X^{}$ ($X=Z, b, \gamma$ and $h$), where $b$ represents 
the bottom quark. The following input values are used~\cite{PDG}; 
\begin{align}
\label{input}
&m_Z=91.1875{\rm GeV},\ \ G_F=1.16639\times10^{-5}{\rm GeV^{-2}}, 
\\ \notag
&\alpha_\textrm{EM}^{-1}=137.035989,\ \  \Delta \alpha_\textrm{EM}=0.06635, 
\\ \notag
&m_t=173.07{\rm GeV},\ \ m_b=4.66{\rm GeV},\ \  m_c=1.275{\rm GeV},\ \ m_{\tau}=1.77684{\rm GeV}, 
\\ \notag
&m_h=125{\rm GeV}. 
\end{align}

We investigate the following regions for masses of the inert scalar bosons in each scenario,
\begin{align}
  &\textrm{Scenario-A}\quad 100\textrm{ GeV}\leq m_{H^\pm} \leq 1000\textrm{ GeV}, \label{range-A}\\
  &\textrm{Scenario-B}\quad 500\textrm{ GeV}\leq m_{H^\pm} \leq 1000\textrm{ GeV}, \label{range-B2}\\
  &\textrm{Scenario-C}\quad 100\textrm{ GeV}\leq m_{\Phi} \leq 1000\textrm{ GeV},\
   0 \leq \mu_2^2 \leq (2000 \textrm{ GeV})^2, \label{range-C} 
\end{align}
where $\Phi$ represents $H$, $A$ and $H^\pm$. 
For all scenarios, we set $\lambda_2$ to be 1. 
The other parameters for Scenario-A and Scenario-B are shown in TABLE~\ref{tab}.
For Scenario-A and Scenario-C, we take the mass of inert scalar boson from $100$ GeV to 1 TeV except that for the dark matter candidate $A$, taking into acount the bound on $m_{H^{\pm}}$ from the LEP experiment~\cite{LEP-II-H, LEP-II-Hp}.  

 \begin{figure}
   \centering
\includegraphics[width=5.5cm,angle=270]{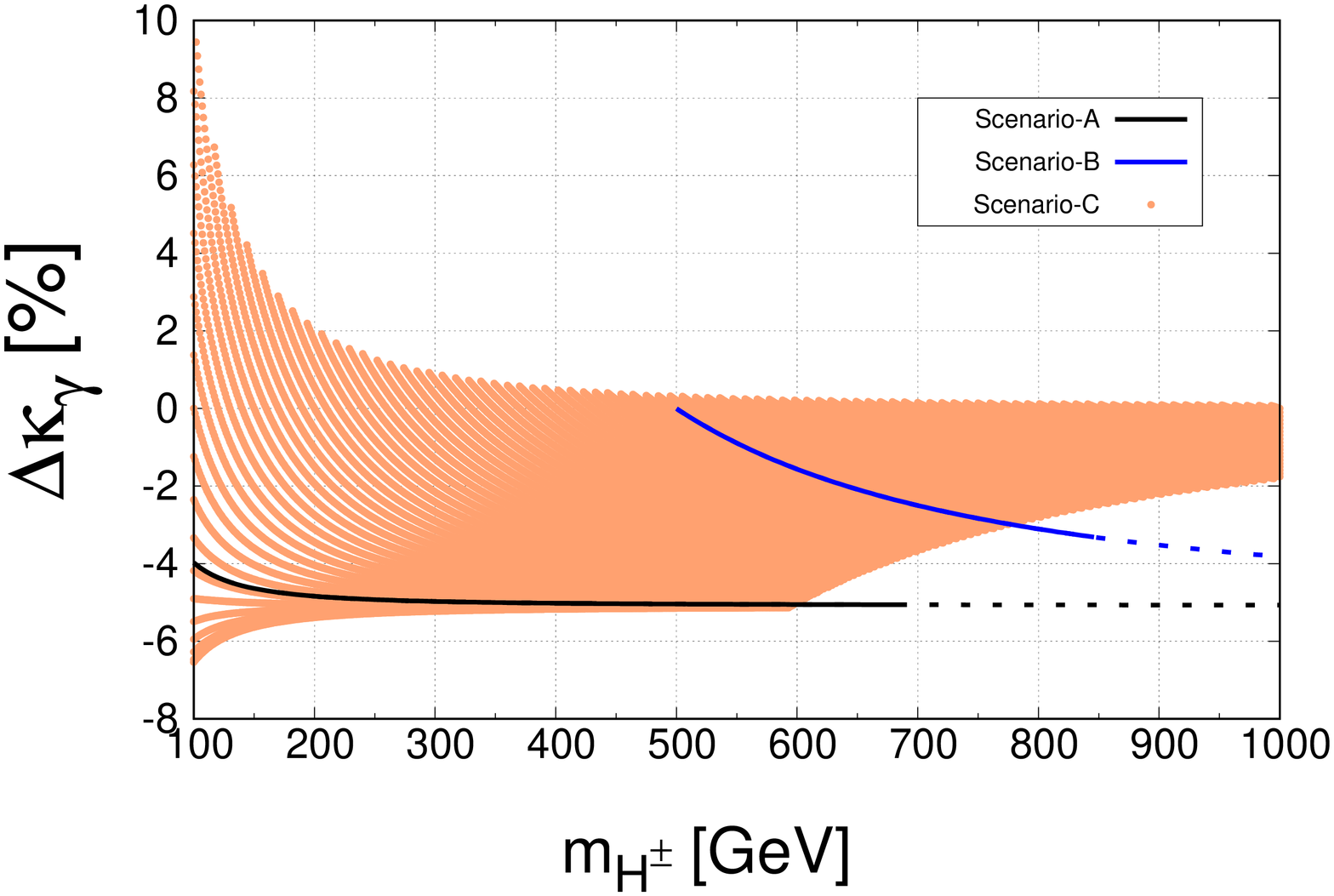}\vspace{1mm}    \hspace{1mm}    
\includegraphics[width=5.5cm,angle=270]{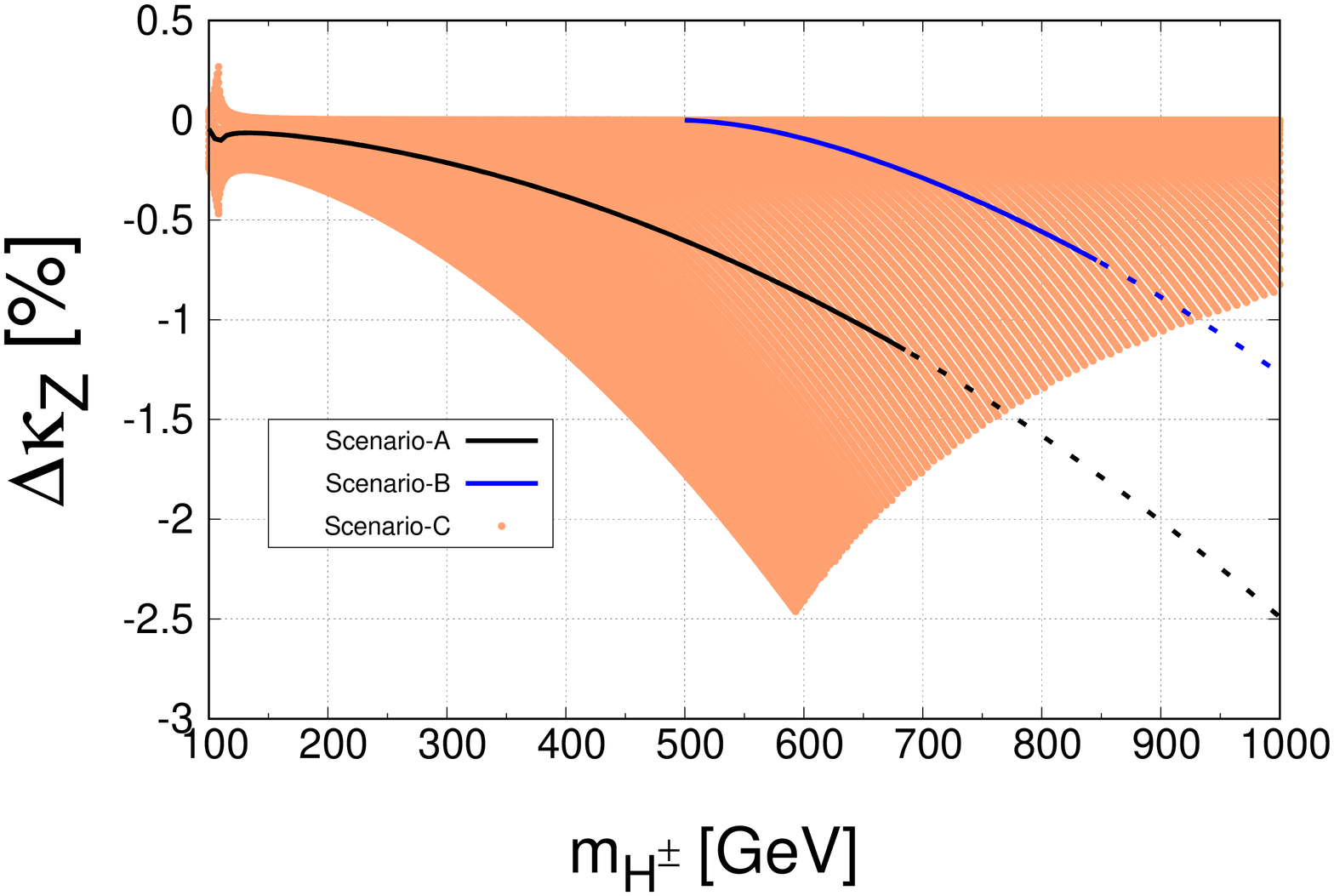}	\vspace{1mm}    
\includegraphics[width=5.5cm,angle=270]{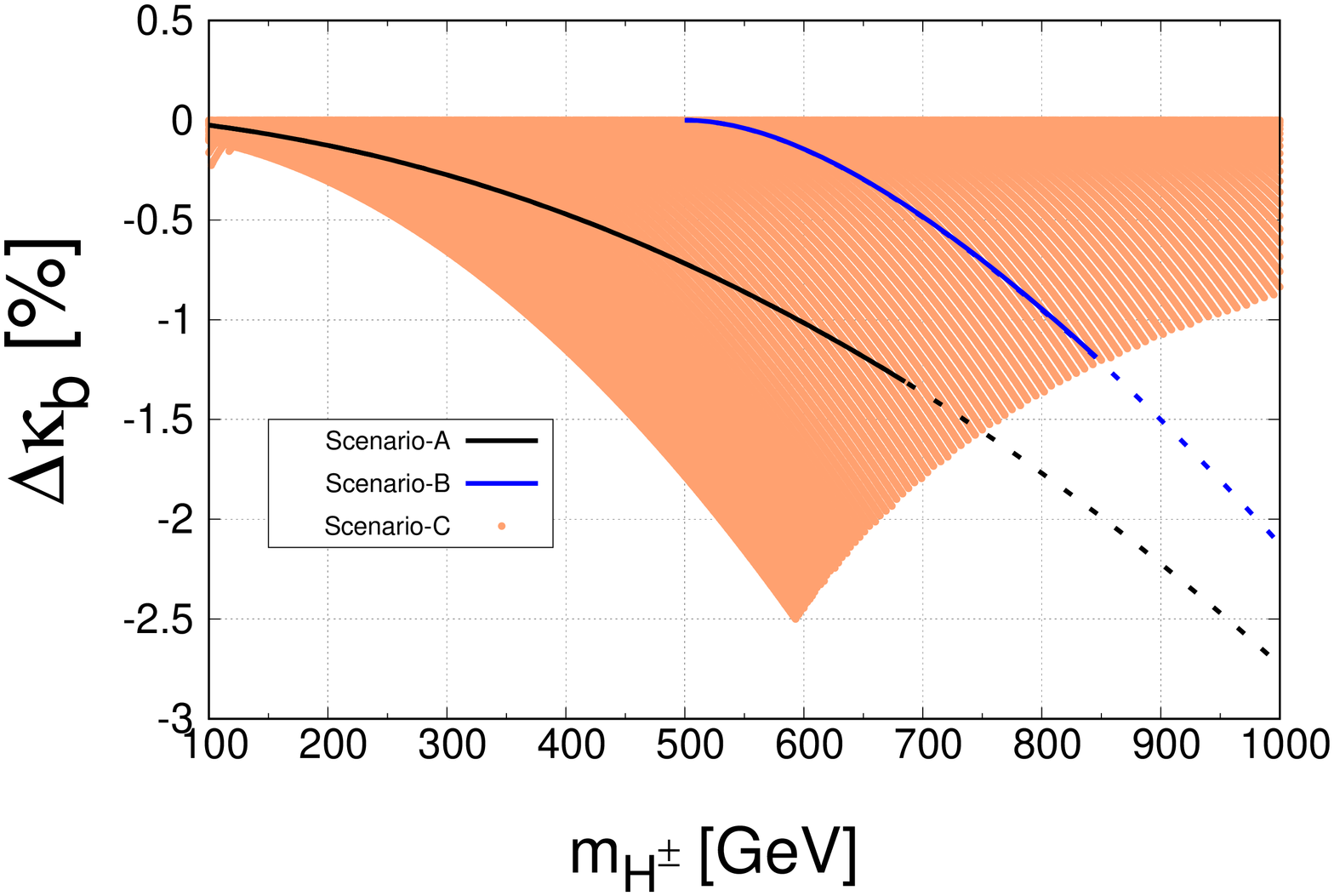}	\vspace{1mm}  \hspace{1mm}    
\includegraphics[width=5.5cm,angle=270]{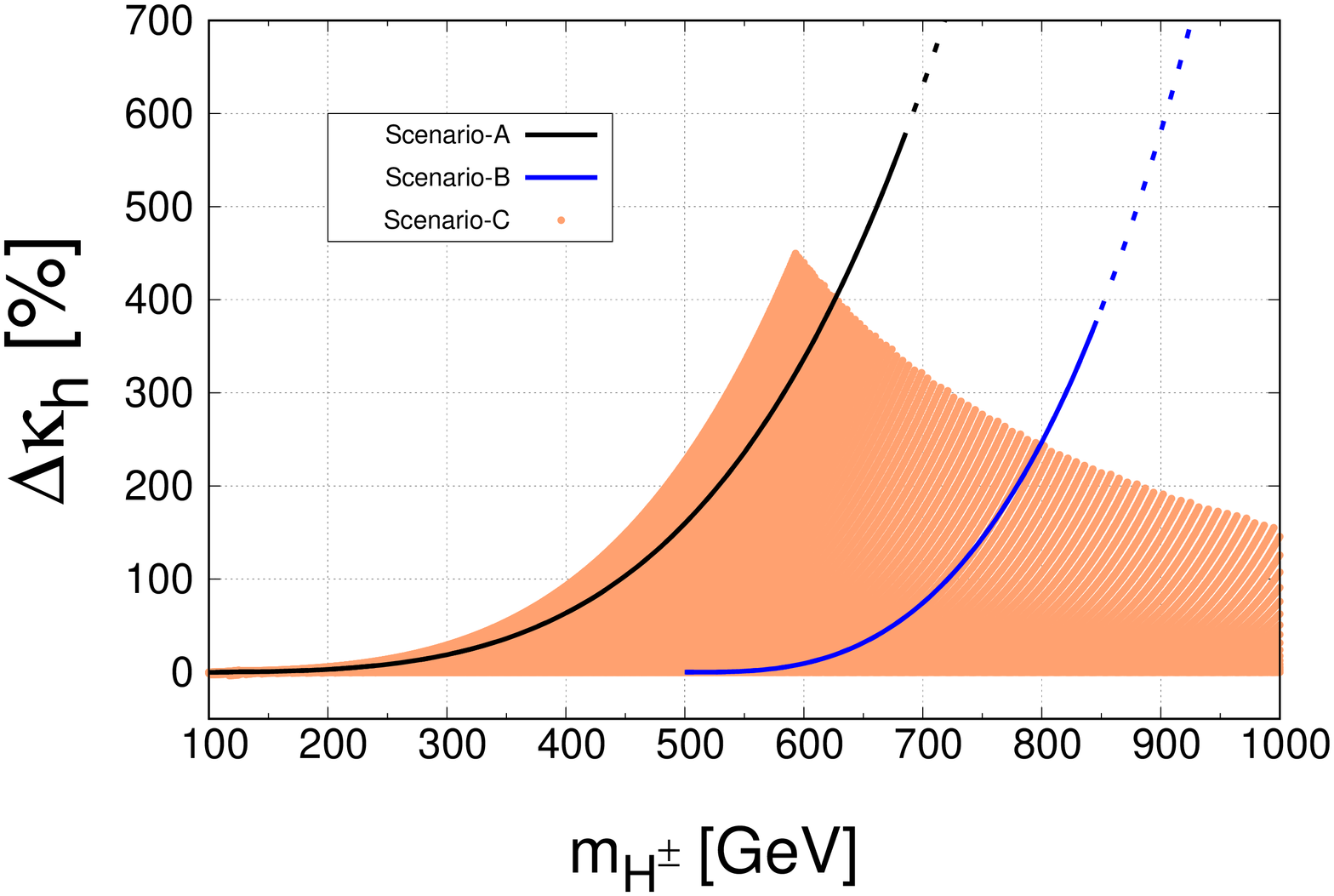}	\vspace{1mm}
\caption{
Predicted regions of $\Delta \hat{\kappa}_\gamma^{}$, $\Delta \hat{\kappa}_{Z}^{}$, $\Delta \hat{\kappa}_{b}^{}$ and $\Delta\hat{\kappa}_h^{}$ as a function of $m_{H^\pm}$ under the constraint from vacuum stability and perturbative unitarity.
In each figure, the regions in Scenario-A (Black), Scenario-B (Blue) and Scenario-C (Orange) are shown. 
For Scenario-A (Scenario-B), $\mu_2$ is set to be 61.50 GeV (499.9GeV), while  
for the results of Scenario-C $\mu_2$ is scanned from 0 to 2~TeV.
The regions of the dotted lines of Scenario-A and Scenario-B are excluded by the perturbative unitarity bound.}
\label{Fig:mHp_dk}
 \end{figure}

 In Fig.~\ref{Fig:mHp_dk}, we show the results for $\Delta\hat{\kappa}_X^{}$ ($X = \gamma, Z, b, h$) of Scenario-A (Black), Scenario-B (Blue) and Scenario-C (Orange) as a function of $m_{H^{\pm}}^{}$ under the 
 theorectical constraints from perturbative unitarity and vacuum stability.

 First, we discuss the result of $\Delta\hat{\kappa}_\gamma^{}$ shown in the top panel of Fig.~\ref{Fig:mHp_dk}. 
 In Scenario-A, the value of $\Delta\kappa_\gamma^{}$ keeps to be 4 - 5 \% without depending on $m_{H^\pm}^{}$.
 We can understand from Eq.~(\ref{eq:appro_dkg}) the reason why the $H^\pm$ loop corrections is usually non-zero value. 
 Since the value of $\mu_2^2$ is fixed to be about (61 GeV)$^2$, the non-decoupling effect is given in the region for $m_{H^\pm}^{} > 100$ GeV.
 In Scenario-B, we can see that $\Delta\kappa_\gamma^{}$ is equal to 0 when $H^\pm$ and $A$ are degenerated, 
 because of $\mu_2^2 \simeq m_A^2$.
 The $h\gamma\gamma$ coupling can deviate from the SM prediction by maximally about $-3.2$ \%.  
 For predictions of Scenario-A and Scenario-B, parts of the dashed curves are excluded by the perturbative unitarity bound.
 $\kappa_\gamma^{}$ is expected to be measured with 2-5 \% (1$\sigma$) accuracy at the HL-LHC~\cite{HWGR, HL-LHC}  
 so that these scenarios may be tested by detecting such deviations in the $h\gamma\gamma$ coupling.
 In particular, we can exclude Scenario-A if a significant deviation is not detected. 
In Scenario-C, $\Delta\hat{\kappa}_\gamma^{}$ is expected to be from about $-7$\% to about 9\% in small mass region. 
$\Delta\kappa_\gamma^{}>0$ is caused by the negative sign of $\lambda_3^{}$ which determine the form of the $hH^+H^-$ coupling as
 $\lambda_{hH^+H^-}^{}= -\lambda_3^{}v$.
 For larger region of $m_{H^\pm}^{}$ than about 600GeV, smaller values of $\Delta\hat{\kappa}_\gamma^{}$ are excluded by the unitarity bound so that
 $\Delta\hat{\kappa}_\gamma^{}$ approaches to 0 in the large mass limit. 

 From the top right panel and the down left panel of Fig.~\ref{Fig:mHp_dk}, we can find that the behavior of $\Delta\kappa_Z^{}$ is similar to that of $\Delta\kappa_b^{}$ in all scenarios.  
 In Scenario-A and Scenario-B, 
 the deviations in the $hZZ$ coupling is suppressed compared to that of the $h b\bar{b}$ coupling, 
 because the cancelation happens between the 1PI diagram contributions 
 to $\Gamma_{hZZ}^{1\PI}$ and those to $\delta Z_h^{}$ in $\delta \Gamma_{hZZ}^1$
 in Eqs.~(\ref{hVV_sm}) and (\ref{hVV_1_ana}).  
 We can see that magnitudes of deviations on the $hZZ$ and the $hb\bar{b}$ couplings enhance as $m_{H^\pm}^{}$ becomes large. 
 In Scenario-A and Scenario-B,  
 maximal values of $|\Delta \kappa_{Z}|$ ($|\Delta\kappa_b^{}|$) are about 1.2\% (1.3\%) and 
 0.8\% (1.2\%) at $m_{H^\pm}=680$ GeV and $850$ GeV which are given by the perturbative unitarity bound, respectively. 
 In Scenario-C, the maximal value of the one-loop corrections to the $hZZ$ and $hb\bar{b}$ couplings are about $-2.5$ \%
 which is realized in $m_{H^\pm}^{} \simeq 600$ GeV. 
 $|\Delta\kappa_{Z(b)}^{}|$ grows with the speed $m_{\Phi}^2$ to $m_{\Phi}^{}\simeq 600$ GeV.
 In the region with $m_{\Phi}^{}\gtrsim 600$ GeV, the unitarity bound reduces deviations in the $hZZ$ and $h\bar{b}b$ couplings. 
In these figures, a bump can be seen at around $m_{H^\pm} = (m_h+m_Z)/2 \simeq 110$ GeV. This can be 
understood as a threshold effect. The direction of the threshold effect depends on the sign of $\lambda_{h\Phi\Phi}^{}$ 
$(=-\lambda_3 v)$, which can be both positive and negative under the vacuum stability bound in Eq.~(\ref{eq:vs}). 
Except for these regions of the threshold enhancement, $\Delta\hat{\kappa}_Z$ and $\Delta\hat{\kappa}_b$ 
do not deviate to the positive direction from the SM prediction.   
This is consistent with our approximate analytic formula in Eq.~(\ref{eq:dkz_appro1}).
These results, however, are different from those in Ref.~\cite{idm_hhh_ref}, where about $+2.2$\% of the positive deviation 
from the SM prediction is seen in the $hZZ$ coupling. 

 The down right panel shows $\Delta\hat{\kappa}_h^{}$ in all scenarios. 
We can find that extremely significant deviations can appear in the $hhh$ coupling in all scenarios.  
$\Delta\hat{\kappa}_h^{}$ can be typically 100\%, and maximally about 600 \%,  400 \% and 450 \% in Scenario-A, Scenario-B 
and Scenario-C, respectively, 
under the constraints from perturbative unitarity and vacuum stability.
Each value of $m_{H^\pm}^{}$ where the maximal value of $\Delta\hat{\kappa}_h^{}$ is realized corresponds to 
those of $\Delta\hat{\kappa}_Z^{}$ and $\Delta\hat{\kappa}_b^{}$.  
 Such a large deviation can be realized by non-decoupling loop effects of the additional scalar bosons $\Phi$ 
 with quartic power of $m_{\Phi}^{}$, which was known in the other extended Higgs sectors~\cite{KKOSY, KOSY, HTM_reno2}.  
 One of the physics cases in which these non-decoupling effects appear is the case with the strongly first order phase transition 
 which is required for a successful scenario of electroweak baryogenesis~\cite{EWPT,KOS}.  
  The triple Higgs boson coupling is expected to be 
 measured with the 54 \% accuracy at the HL-LHC~\cite{HL-LHC_hhh}. 
 Moreover, it can be measured by O(10) \% at the ILC with $\sqrt{s}=1$ TeV (ILC1TeV)~\cite{ILC_TDR,white_paper,CLIC}.
 If a significant deviation in the $hhh$ coupling is detected in the future at these colliders,  
 we can extract information of the mass of inert scalar bosons indirectly. 

 We mention the direct rearch of the inert scalar bosons at future collider experiments. 
 According to several current studies, in Scenario-A, the direct discovery reach of $H$ may be about 200 GeV at the LHC with $\sqrt{s}=14$ TeV and the integrated luminosity is 300 fb$^{-1}$.
 The discovery rearch may be up to 300 GeV at the HL-LHC with $\sqrt{s}=14$ TeV and the integrated luminosity to be 3000 fb$^{-1}$.
Therefore, such a smaller mass region is expected to be tested by direct searches of inert scalar particles at future collider experiments. 
On the other hand, for the case where inert scalar particles are too heavy to be directly detected,
the indirect test by using precision measurements of the Higgs boson couplings can be a good approach to detect the IDM 
complementarily.

 \begin{figure}
\centering
\includegraphics[width=5.5cm,angle=270]{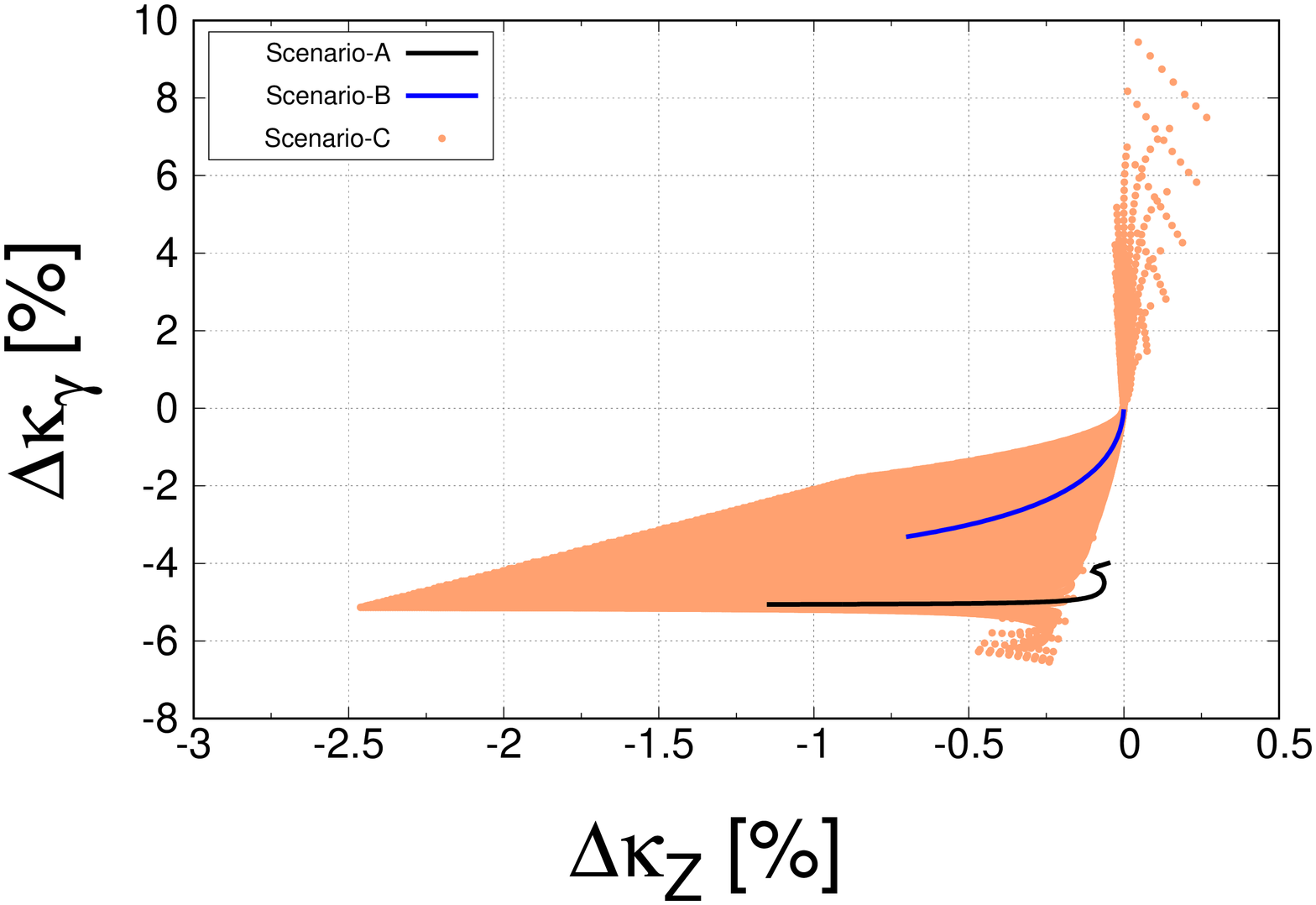}\vspace{10mm}
    \includegraphics[width=5.5cm,angle=270]{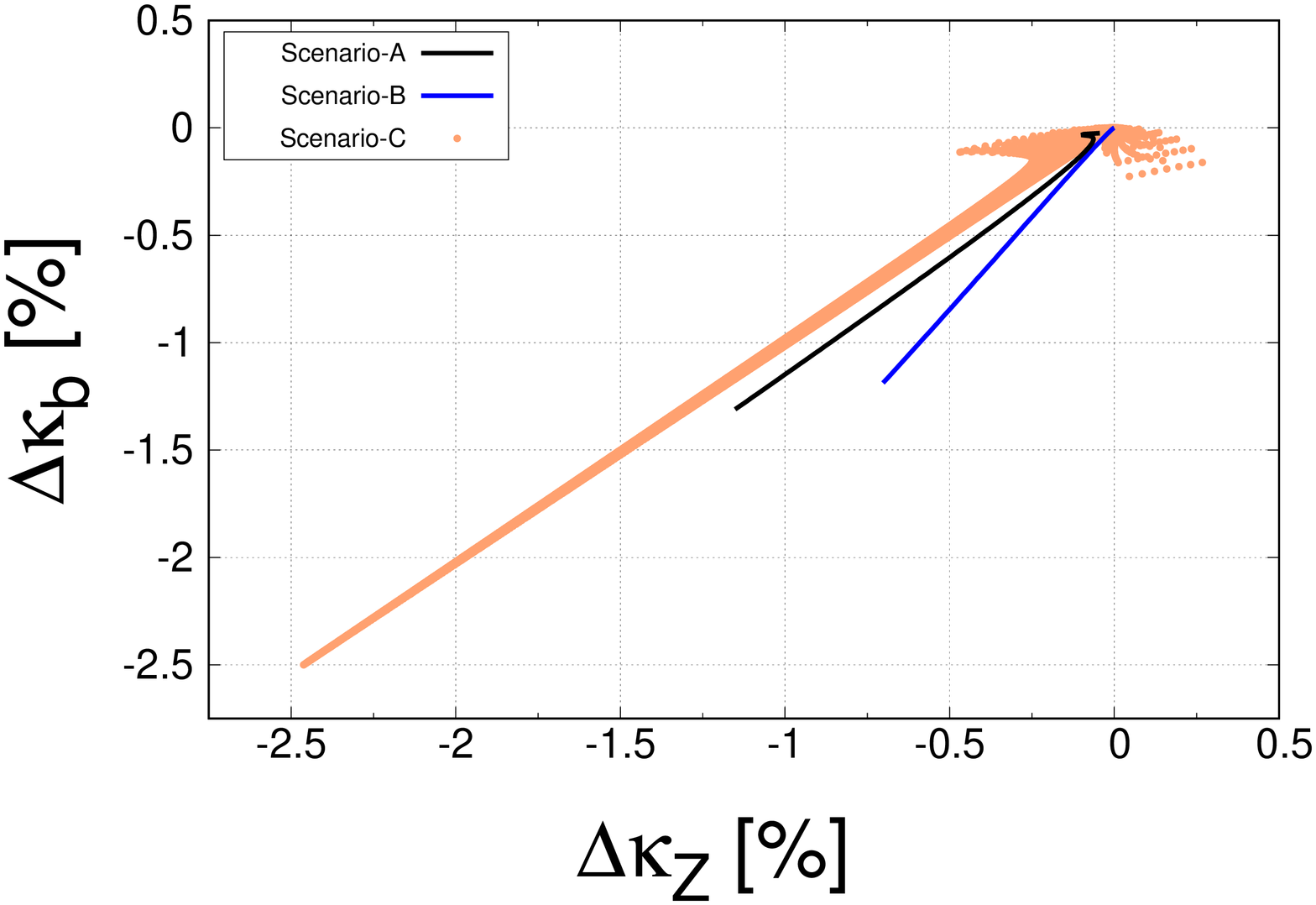}	\vspace{1mm}    
\includegraphics[width=5.5cm,angle=270]{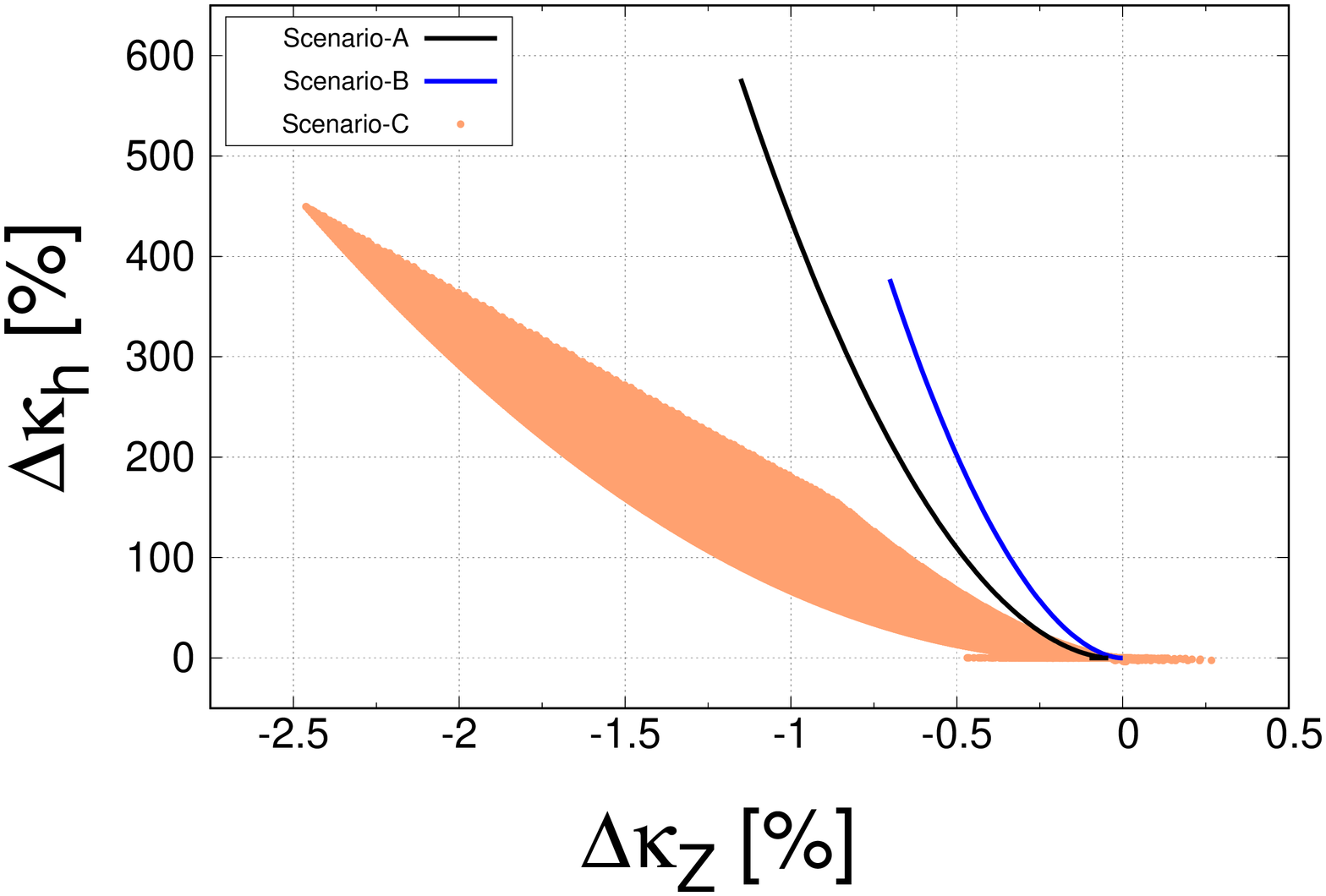}	\vspace{1mm}
 \caption{
 Predicted regions of Scenario-A (Black), Scenario-B (Blue) and Scenario-C (Orange) 
 on the $\Delta \kappa_{Z}$- $\Delta \kappa_{\gamma}$ plane, 
 the $\Delta \kappa_{Z}$- $\Delta \kappa_{b}$ plane and the $\Delta \kappa_{Z}$- $\Delta \kappa_{h}$ plane 
 under the constraint from vacuum stability and perturbative unitarity. 
 For Scenario-A (Scenario-B), $\mu_2$ is set to be 61.50 GeV (499.9GeV), while  
for the results of Scenario-C $\mu_2$ is scanned from 0 to 2~TeV. }
 \label{Fig:corr}
 \end{figure}

Finally, we discuss patterns of deviations in the $h\gamma\gamma$, $hZZ$, $hb\bar{b}$ and $hhh$ vertices in Scenario-A, 
Scenario-B and Scenario-C. 
If the Higgs sector is extended from the minimal one like the IDM, 
the Higgs boson couplings can deviate from the predictions in the SM. 
The pattern of the deviations in various Higgs boson couplings 
largely depend on the structure of extended Higgs sectors.
Therefore, such a pattern  in each model can be useful to discriminate the model from 
the other models with extended Higgs sectors when they are detected at future colliders.

In Fig.~\ref{Fig:corr}, we show correlations among $\Delta\hat{\kappa}_\gamma^{}$, $\Delta\hat{\kappa}_Z^{}$, $\Delta\hat{\kappa}_b^{}$ and $\Delta\hat{\kappa}_h^{}$ in Scenario-A, Scenario-B and Scenario-C.  
We survey parameter regions given in Eqs.~(58)-(60) under the constraints from purturbative unitarity and vacuum stability in each scenario. 
Definition of colors for the predicted regions of these scenarios are the same as those in Fig.~\ref{Fig:mHp_dk}. 
What we can learn from these results? 
Suppose that in the future the deviation in $\kappa_Z$ is detected and its central value is about -1\%, 
we can discriminate the model from the SM by measuring $\kappa_\gamma$ if $\kappa_\gamma$ 
deviates from the SM predictions in the negative  direction, 
and we can exclude Scenario-B. In addition, by measuring the $hhh$ coupling, we can separate 
Scenario-A and Scenario-C. 

 \section{Conclusion}
 
  We have evaluated radiative corrections to the Higgs boson couplings in the IDM.  
  The one-loop contributions to the $hVV$, $hff$ and $hhh$ couplings have been calculated in the on-shell scheme. 
   We have evaluated the scaling factors for these couplings and have investigated 
   how the one-loop corrected Higgs boson couplings can deviate from the SM predictions under the
  constraints from perturbative unitarity and vacuum stability in the scenarios where the model can explain current dark matter data. 
  When the mass of the dark matter is slightly above $m_h/2$ (Scenario-A), future direct searches such as XENON-1T should 
  be able to survey the model to some extent but there will be still a region  
  which cannot be excluded where $\lambda_A^{}\simeq \mathcal{O}(10^{-4})$ or less.  
  We have found that  even in such a case 
  we may be able to investigate the model  by future collider experiments 
  by either the direct search of heavier inert particles or future precision measurements of the Higgs boson couplings.   
  In addition, for relatively heavy dark matter scenarios such as Scenario-B, the Higgs couplings can receive 
  relatively large corrections so that the future precision measurement can also be useful to test these cases. 
  In conclusion, in addition to direct/indirect dark matter experiments and the direct search experiments at colliders,  
  future precision measurements for the Higgs boson couplings at the HL-LHC or the ILC can play an important role 
  to indirectly test the dark matter scenarios in the IDM as a complementary tool. \\

\noindent
{\it Acknowledgments}

S.K. was supported in part by Grant-in-Aid for Scientific Research, The Ministry of Education, Culture, 
Sports, Science and Technology (MEXT), No. 23104006, and Grant H2020-MSCA-RISE-2014 No. 645722 (Non-Minimal Higgs). 
M.K. was supported in part by JSPS, No. K25$\cdot$10031.

\newpage 

\appendix
\section{1PI diagram contributions to the Higgs boson coupling} 

We here give all 1PI diagram contributions to the renormalized $hVV$, $hf\bar{f}$ and $hhh$ couplings in the IDM. 
Calculations for the renormalized Higgs boson couplings are performed in 't Hooft-Feynman gauge so that the masses of Numbu-Goldstone bosons $m_{G^\pm}^{}$ and $m_{G^0}^{}$
and those of Fadeev-Popov ghosts $m_{c^\pm}^{}$, $m_{c_Z^{}}^{}$ and $m_{c_\gamma^{}}^{}$ are the same as corresponding masses of the gauge bosons. 
The tree level scalar coupling constants defined as,
\begin{align}
  \mathcal{L} =  \lambda_{\phi_1^{}\phi_2^{}\phi_3^{}\phi_4^{}}^{} \phi_1^{}\phi_2^{}\phi_3^{}\phi_4^{}  +\cdots,
\end{align}
are given by, 
\begin{align}
	\lambda_{hhHH}=-\frac{m_H^2-\mu^2_2}{2v^2}\ , \ 
	\lambda_{hhAA}=-\frac{m_A^2-\mu^2_2}{2v^2}\ , \ 
	\lambda_{hhH^{+}H^{-}}=-\frac{m^2_{H^+}-\mu^2_2}{v}, \\
	\lambda_{hhhh}=-\frac{m^2_h}{8v^2} \ , \ 
	\lambda_{hhG^0G^0}=-\frac{m^2_h}{4v^2} \ , \ 
	\lambda_{hhG^+G^-}=-\frac{m^2_h}{2v^2}. \ \ \ \ \ \ \ \ \ \ \ \ \ \ 
\end{align}

We show one-loop contributions of 1PI diagrams to one-, two- and three-point functions by using Passarino-Veltman functions~\cite{PV_func} 
according to the notation in Ref.~\cite{Hagiwara}.
We write 1PI diagram contributions to each function separately for fermion loop contributions and boson loop contributions which are
expressed by index $F$ and $B$, respectively. 

One-loop contributions to the tadpole state $\Gamma_{h,(F,B)}^{1\PI}$ are given by,
\begin{align}
\label{1Ff_h}
	16\pi^2 \Gamma^{1\PI}_{h,F} &= -\sum_{f} N_c^f \frac{4m_f^2}{v} A(m_f^2),  \\ \notag
\label{1FB_h}
	16\pi^2  \Gamma_{h,B}^{1\PI}&=-\lambda_{hH^+H^-}A(m_{H^{\pm}})-\lambda_{hAA}A(m_{A})-\lambda_{hHH}A(m_{H})  \\ \notag
	&-3\lambda_{hhh}A(m_h)-\lambda_{hG^+G^-}A(m_{G^{\pm}})
	-\lambda_{hG^0G^0}A(m_{G^0})  \notag \\
	&+3gm_WA(m_W)+\frac{3}{2}g_Zm_ZA(m_Z)-2gm_W^3-g_Zm_Z^3. 
\end{align}
where $N_c^f$ indicates the color number of each particle. 

One-loop corrections to transverse parts of two-point functions of gauge bosons are calculated as,
\begin{align}
\label{2Ff_gamgam}
16\pi^2\Pi_{\gamma\gamma, T}^{\rm 1PI}(p^2)_F&=\sum_f8e^2Q_f^2N_c^fp^2B_3(p^2;m_f,m_f), \\
\label{2Ff_Zgam}
16\pi^2\Pi_{Z\gamma, T}^{\rm 1PI}(p^2)_F&=\sum_f eg_ZN_c^f\Big[2p^2(2I_fQ_f-4s_W^2Q_f^2) \Big]B_3(p^2;m_f,m_f),  \\
\label{2Ff_ZZ}
	16\pi^2 \Pi^{\rm 1PI}_{ZZ, T}(p^2)_F &= \sum_{f}N^f_c g_Z^2
	\Bigg[2p^2\left(2 I^2_f -4I_f Q_fs^2_W+4Q_f^2s_W^4\right)B_3 
	 -2I_f^2m_f^2B_0\Bigg](p^2;m_f,m_f), \\
\label{2Ff_WW}
	16\pi^2 \Pi^{\rm 1PI}_{WW, T}(p^2)_F &= \sum_{f,f^{\prime}} N^f_c g^2
	\left\{ 2p^2B_3-B_4\right\}(p^2;m_f,m_{f^{\prime}}), 
\end{align}
\begin{align}
\label{2FB_gamgam}
16\pi^2\Pi_{\gamma\gamma, T}^{\rm 1PI}(p^2)_B&=e^2B_5(p^2;m_{H^{\pm}},m_{H^{\pm}})-e^2p^2\Big[12B_3+5B_0+\frac{2}{3} \Big](p^2;m_W,m_W), \\ 
\label{2FB_Zgam}
16\pi^2\Pi_{Z\gamma, T}^{\rm 1PI}(p^2)_B&=\frac{eg_Z}{2}B_5(p^2;m_{H^{\pm}},m_{H^{\pm}})\notag\\
&-eg_Zp^2\Big[10B_3+\frac{11}{2}B_0+\frac{2}{3} \Big](p^2;m_W,m_W)-16\pi^2\frac{s_W}{c_W}\Pi_{\gamma\gamma}^{\rm 1PI}(p^2)_B,
\\ \notag
\label{2FB_ZZ}
16\pi^2 \Pi_{ZZ, T}^{\rm1PI}(p^2)_B&=
\frac{g_Z^2}{4}\left\{B_5(p^2;m_{H^{\pm}},m_{H^{\pm}})+B_5(p^2;m_H,m_A)\right\} \\ \notag
	   &+g_Z^2\Big(m_Z^2B_0+\frac{1}{4}B_5\Big)(p^2;m_h,m_Z) \\ \notag
	   &+g_Z^2\Big\{\Big(2m_W^2-\frac{23}{4}p^2\Big)B_0-9p^2B_3 
	   -\frac{2}{3}p^2\Big\}(p^2;m_W,m_W) \\ 
	   &+16\pi^2 \Big\{-\frac{2s_W}{c_W}\Pi_{Z\gamma}^{\rm 1PI}(p^2)_B
	   -\frac{s_W^2}{c_W^2}\Pi_{\gamma\gamma}^{\rm 1PI}(p^2)_B \Big\},
	   \\ \notag
\label{2FB_WW}
	   16\pi^2  \Pi^{1\PI}_{WW, T}(p^2)_B&=
           	\frac{g^2}{4}\left\{B_5(p^2;m_H,m_{H^{\pm}})+B_5(p^2;m_A,m_{H^{\pm}})\right\}  \\\notag
	&+g^2\Big(m_W^2B_0+\frac{1}{4}B_5\Big)(p^2;m_h,m_W) \\ \notag
	 &+g^2\Big\{\Big(\frac{1}{4}+2c_W^2\Big)B_5+(m_W^2-4s_W^2m_W^2+m_Z^2-8p^2c_W^2)
	   B_0\Big\}(p^2;m_Z,m_W) \\  
	  &+2s_W^2\big\{B_5+(2m_W^2-4p^2)B_0\Big\}(p^2;m_{\gamma},m_W)
	  -\frac{2}{3}g^2p^2,
\end{align}
where $g_Z^{}=g/c_W^{}$.

The vector, axial vector and scalar parts ($\Pi_{ff,V}^{1\PI}$, $\Pi_{ff,A}^{1\PI}$ and $\Pi_{ff,S}^{1\PI}$) 
of one-loop contributions to two-point functions of a fermion field is identical to those of the SM, 
because the inert scalar particles do not couple to fermion directly.
$\Pi_{ff,V}^{1\PI}$, $\Pi_{ff,A}^{1\PI}$ and $\Pi_{ff,S}^{1\PI}$ are calculated as 
\begin{align}
\notag
\label{2F_ff_V}
16\pi^2\Pi_{ff,V}^{1\PI}(p^2)&=-\frac{g^2}{4}(2B_1+1)(p^2;m_{f^{\prime}},m_W)
-g_Z^2(v_f^2+a_f^2)(2B_1+1)(p^2;m_{f^{}},m_Z) \\ \notag
&-e^2Q_f^2(2B_1+1)(p^2;m_f,m_{\gamma})-\frac{m_f^2}{v^2}B_1(p^2;m_f,m_h) \\ 
&-\frac{m_f^2}{v^2}B_1(p^2;m_f,m_{G^0})
-\frac{m_f^2+m_{f^{\prime}}^2}{v^2}B_1(p^2;m_{f^{\prime}},m_{G^{\pm}}), \\ \notag
16\pi^2\Pi_{ff,A}^{1\PI}(p^2)&=-\frac{g^2}{4}(2B_1+1)(p^2;m_{f^{\prime}},m_W)
-2g_Z^2v_fa_f(2B_1+1)(p^2;m_{f^{}},m_Z) \\ 
&+\frac{m_f^2-m_{f^{\prime}}^2}{v^2}B_1(p^2;m_{f^{\prime}},m_{G^{\pm}}), \\ \notag
\label{2F_ff_S}
16\pi^2\Pi_{ff,S}^{1\PI}(p^2)&=-2g_Z^2(v_f^2-a_f^2)(2B_0-1)(p^2;m_f,m_Z)
-2e^2Q_f^2(2B_0-1)(p^2;m_f,m_{\gamma}) \\ 
&-2\frac{m^2_{f^{\prime}}}{v^2}B_0(p^2;m_{f^{\prime}},m_{G^{\pm}})
+\frac{m^2_{f^{}}}{v^2}B_0(p^2;m_{f^{}},m_{h})
-\frac{m^2_{f^{}}}{v^2}B_0(p^2;m_{f^{}},m_{G^0}),
\end{align}
where $v_f=I_f/2-s_W^2Q_f$ and $a_f=I_f/2$.

1PI diagram contributions to the Higgs boson are given by   
\begin{align}
  \label{2Ff_hh}
	16\pi^2 \Pi^{1\PI}_{hh}(p^2)_{F} &= -\sum_{f} N_c^f \frac{m_f^2}{v^2} \left\{ 4A(m_f)+(-2p^2+8m^2_f)B_0(p^2;m_f,m_f)\right\}, \\
\notag
\label{2FB_hh}
	16\pi ^2 \Gamma_{hh}^{1\PI}(p^2)_B&=
        	\lambda_{hH^+H^-}^2B_0(p^2;m_{H^{\pm}},m_{H^{\pm}}) 
	\\ \notag
	&+2\lambda_{hHH}^2B_0(p^2;m_{H},m_{H}) +2\lambda_{hhA}^2B_0(p^2;m_{A},m_{A})
	\\\notag
	 &-2\lambda_{hhH^+H^-}A(m_{H^{\pm}})-2\lambda_{hhHH}A(m_{H})-2\lambda_{hhAA}A(m_{A}) \\\notag
        &+18\lambda^2_{hhh}B_0(p^2;m_h,m_h)
	+\lambda^2_{hG^+G^-}B_0(p^2;m_{G^{\pm}},m_{G^{\pm}}) \\ \nonumber
	&+2\lambda^2_{hG^0G^0}B_0(p^2;m_{G^0},m_{G^0})  \\ \notag
	&-12\lambda_{hhhh}A(m_h)
	-2\lambda_{hhG^+G^-}A(m_{G^{\pm}}) 
	-2\lambda_{hhG^0G^0}A(m_{G^0}) \\ \notag
	&+g^2(3m_W^2-p^2)B_0(p^2;m_W,m_W)+\frac{g^2}{2}3A(m_W)-3g^2m_W^2 \\ 
	&+\frac{g_Z^2}{2}(3m_Z^2-p^2)B_0(p^2;m_Z,m_Z)+\frac{g_Z^2}{4}3A(m_Z)
	-\frac{3}{2}g_Z^2m_Z^2.
\end{align}

Next, we give analytic expressions for the 1PI diagram contributions to the $hVV$, $hf\bar{f}$ and $hhh$.
We use the simplified form for the three point function of the Passarino-Veltman function as $C_i(X,Y,Z)\equiv C_i^{}(p_1^2,p_2^2,q^2;m_X^{},m_Y^{},m_Z^{})$.

The 1PI diagram contributions to the $hhh$ coupling are given by 
\begin{align}
\notag
\label{3Ff_hhh}
 16\pi^2 \Gamma^{1\PI}_{hhh}(p^2_1,p^2_2,q^2)_F &=  -\sum_{f} N_c^f
\frac{8m^4_f}{v^3}\{ B_0(p^2_1,m_f,m_f)+B_0(p^2_2,m_f,m_f) \\ \nonumber 
&+B_0(q^2,m_f,m_f)
+(4m_f^2-q^2+p_1 \cdot p_2)C_0(m_f,m_f,m_f)\}, \\
\end{align}
\begin{align}
\notag
\label{3FB_hhh}
16\pi^2  &\Gamma_{hhh}^{1\PI}(p^2_1,p^2_2,q^2)_B \\ \notag
&=
	2\lambda_{hH^+H^-}\lambda_{hhH^+H^-}\Big\{
	B_0(p^2_1;m_{H^+},m_{H^+})+B_0(p^2_2;m_{H^+},m_{H^+})+B_0(q^2;m_{H^+},m_{H^+})
	\Big\} \\ \notag
	&+4\lambda_{hAA}\lambda_{hhAA}\Big\{
	B_0(p^2_1;m_{A},m_{A})+B_0(p^2_2;m_{A},m_{A})+B_0(q^2;m_{A},m_{A})
	\Big\} \\ \notag
	&+4\lambda_{hHH}\lambda_{hhHH}\Big\{
	B_0(p^2_1;m_{H},m_{H})+B_0(p^2_2;m_{H},m_{H})+B_0(q^2;m_{H},m_{H})
	\Big\} \\ \notag
	&-2\lambda_{hH^+H^-}^3C_0(m_{H^+},m_{H^+},m_{H^+})
	-8\lambda_{hAA}^3C_0(m_{A},m_{A},m_{A})
	-8\lambda_{hHH}^3C_0(m_{H},m_{H},m_{H}) \\\notag
		&+2\lambda_{hG^+G^-}\lambda_{hhG^+G^-}\Big\{
	B_0(p^2_1;m_{G^{\pm}},m_{G^{\pm}})+B_0(p^2_2;m_{G^{\pm}},m_{G^{\pm}})+B_0(q^2;m_{G^{\pm}},m_{G^{\pm}})
	\Big\} \\ \nonumber 
	&+4\lambda_{hG^0G^0}\lambda_{hhG^0G^0}\Big\{
	B_0(p^2_1;m_{G^0},m_{G^0})+B_0(p^2_2;m_{G^0},m_{G^0})+B_0(q^2;m_{G^0},m_{G^0})
	\Big\} \\ \nonumber 
	&+72\lambda_{hhh}\lambda_{hhhh}\Big\{
	B_0(p^2_1;m_{h},m_{h})+B_0(p^2_2;m_{h},m_{h})+B_0(q^2;m_{h},m_{h})
	\Big\} \\ \nonumber 
	&-2\lambda_{hG^+G^-}^3C_0(m_{G^{\pm}},m_{G^{\pm}},m_{G^{\pm}})
	-8\lambda_{hG^0G^0}^3C_0(m_{G^0},m_{G^0},m_{G^0}) 
	-216\lambda_{hhh}^3C_0(m_{h},m_{h},m_{h})
	\\ \nonumber 
	&+2g^3m_W\Big\{ B_0(p^2_1;m_W,m_W)+B_0(p^2_2;m_W,m_W)
	+B_0(q^2;m_W,m_W)-\frac{3}{2}\Big\} \\ \nonumber 
	&+g_Z^3m_Z\Big\{ B_0(p^2_1;m_Z,m_Z)+B_0(p^2_2;m_Z,m_Z)
	+B_0(q^2;m_Z,m_Z)-\frac{3}{2} \Big\} \\ \nonumber 
	&+\frac{g^3}{2}m_W^3\Big\{16C_0(m_W,m_W,m_W)
	-C_0(m_{c^{\pm}},m_{c^{\pm}},m_{c^{\pm}}) \Big\} \\ \nonumber 
	&+\frac{g^3_Z}{4}m_Z^3\Big\{16C_0(m_Z,m_Z,m_Z)
	-C_0(m_{c_Z},m_{c_Z},m_{c_Z}) \Big\} \\ \nonumber 
	&-\frac{g^3}{2}m_WC_{hhh}^{SVV}(G^{\pm},W^{},W^{}) 
 	+\frac{g^2}{2}\lambda_{hG^+G^-}C_{hhh}^{VSS}(W,G^{\pm},G^{\pm}) \\ 
	&-\frac{g_Z^3}{4}m_ZC_{hhh}^{SVV}(G^0,Z,Z) 
	+\frac{g_Z^2}{2}\lambda_{hG^0G^0}C_{hhh}^{VSS}(Z,G^0,G^0),
	\end{align}
where
\begin{align}
\nonumber
&C_{hhh}^{SVV}(X,Y,Z) \equiv \\ \nonumber 
&\ \ \ \  
	 \Big[ p^2_1C_{21}+p^2_2C_{22}+2p_1\cdot p_2C_{23}+4C_{24}-\frac{1}{2} \\ \nonumber 
	&-(2p^2_1+p_1\cdot p_2)C_{11}
	-(2p_1\cdot p_2+p^2_2)C_{12}+(p^2_1+p_1\cdot p_2)C_0
	 \Big](X,Y,Z), \\ \nonumber 
	&+ \Big[ p^2_1C_{21}+p^2_2C_{22}+2p_1\cdot p_2C_{23}+4C_{24}-\frac{1}{2} \\ 
	&+(3p^2_1-p_1\cdot p_2)C_{11}
	+(3p_1\cdot p_2-p^2_2)C_{12}+2(p_1^2-p_1\cdot p_2)C_0
	 \Big](Z,X,Y) \\ \nonumber
	&+\Big[ p^2_1C_{21}+p^2_2C_{22}+2p_1\cdot p_2C_{23}+4C_{24}-\frac{1}{2} \\ \nonumber 
	&+(3p^2_1+4p_1\cdot p_2)C_{11}
	+(3p_1\cdot p_2+4p^2_2)C_{12}+2(p^2_1+3p_1\cdot p_2+2p^2_2)C_0
	 \Big](Y,Z,X), \\ \nonumber 
	 \end{align}
	 \begin{align}
\nonumber
&C_{hhh}^{VSS}(X,Y,Z) \equiv \\ \nonumber 
 &\ \ \ \ 
	 \Big[ p^2_1C_{21}+p^2_2C_{22}+2p_1\cdot p_2C_{23}+4C_{24}-\frac{1}{2} \\ \nonumber 
	&+2(2p^2_1+p_1\cdot p_2)C_{11}
	+2(2p_1\cdot p_2+p^2_2)C_{12}+4(p^2_1+p_1\cdot p_2)C_0
	 \Big](X,Y,Z) \\ \nonumber 
	 &+
	 \Big[ p^2_1C_{21}+p^2_2C_{22}+2p_1\cdot p_2C_{23}+4C_{24}-\frac{1}{2} \\ \nonumber 
	&+2p_1\cdot p_2C_{11}
	+2p^2_2C_{12}-(p^2_1+2p_1\cdot p_2)C_0
	 \Big](Z,X,Y) \\ \nonumber 
	 &+
	 \Big[ p^2_1C_{21}+p^2_2C_{22}+2p_1\cdot p_2C_{23}+4C_{24}-\frac{1}{2} \\ 
	&-2p_1\cdot p_2C_{11}
	-2p^2_2C_{12}-(p^2_1-p^2_2)C_0
	 \Big](Y,Z,X). 
	 \end{align}

One-loop 1PI contributions to the form factor $\Gamma^{1,1PI}_{hZZ}$ of $hVV$ vertex are expressed by,
\begin{align}
\notag
\label{3Ff_hZZ}
 16\pi^2& \Gamma^{1,1\PI}_{hZZ}(p_1^2,p_2^2,q^2)_F  \\ \nonumber
 &= \sum_{f} N_c^f
\frac{16m^2_fm_Z^2}{v^3}
\Bigg[\left(v^2_f+a^2_f\right)
\Big\{ B_0(p^2_1;m_f,m_f)+B_0(p^2_2;m_f,m_f) \\ \notag
&+2B_0(q^2;m_f,m_f)+(4m_f^2-p^2_1-p_2^2)C_0(m_f,m_f,m_f)
-8C_{24}(m_f,m_f,m_f) \Big\} \\ 
&-(v_f^2-a_f^2)\Big\{B_0(p^2_2,m_f,m_f)+B_0(p^2_1,m_f,m_f)
+(4m_f^2-q^2)C_0(m_f,m_f,m_f) \Big\}\Bigg] ,
\\ \notag
\label{3Ff_hWW}
 16\pi^2 &\Gamma^{1,1\PI}_{{hWW}}(p^2_1,p^2_2,q^2)_F \\ \nonumber
  &= \sum_{f} N_c^f
\frac{4m^2_fm_W^2}{v^3}
\Bigg[\frac{1}{2} B_0(p^2_1;m_f,m_{f^{\prime}})+\frac{1}{2}B_0(p^2_2;m_f,m_{f^{\prime}})+B_0(q^2;m_f,m_f)  \\ 
 &+\frac{1}{2}(2m_f^2+2m_{f^{\prime}}^2-p^2_1-p_2^2)
C_0(m_f,m_{f^{\prime}},m_f)
-4C_{24}(m_f,m_{f^{\prime}},m_f)
\Bigg] +(m_f\leftrightarrow m_{f^{\prime}}),
\end{align}

\begin{align}
\notag
\label{3FB_hZZ}
  16\pi^2  &\Gamma_{hZZ}^{1,1\PI}(p^2_1,p^2_2,q^2)_B \\ \notag
  &=
   \frac{g^2_Z}{4}\Big\{2\lambda_{hH^{+}H^{-}}c^2_{2W}B_0(q^2;m_{H^{\pm}},m_{H^{\pm}})\notag
 \\ \nonumber
 &+2\lambda_{hHH}B_0(q^2;m_H,m_H) +2\lambda_{hAA}B_0(q^2;m_A,m_A)
 \\ \nonumber
 &-8\lambda_{hH^{+}H^{-}}c_{2W}^2C_{24}(m_{H^{\pm}},m_{H^{\pm}},m_{H^{\pm}})
 \\ \nonumber
 &-8\lambda_{hHH}C_{24}(m_H,m_A,m_{H})-8\lambda_{hAA}C_{24}(m_A,m_H,m_{A})\Big\}
 \\\notag
	&+\frac{g_Z^2}{4}
	\Big\{ 2\lambda_{hG^+G^-}c^2_{2W}B_0(q^2;m_{G^{\pm}},m_{G^{\pm}})
	 + 2\lambda_{hG^0G^0}B_0(q^2;m_{G^0},m_{G^0}) \\ \notag
	 &+6\lambda_{hhh}B_0(q^2;m_{h},m_{h}) 
	-8\lambda_{hG^+G^-}c^2_{2W}C_{24}(m_{G^{\pm}},m_{G^{\pm}},m_{G^{\pm}})
	 \\ \notag
	 &-8\lambda_{hG^0G^0}C_{24}(m_{G^0},m_{h},m_{G^0})
	 -24\lambda_{hhh}C_{24}(m_{h},m_{G^0},m_{h}) \Big\}
	  \\ \notag
	  &-g^3m_W\Big\{\frac{s_W^4}{c_W^2}B_0(p_1^2;m_{G^{\pm}},m_W) 
	  +\frac{s_W^4}{c_W^2}B_0(p_2^2;m_{G^{\pm}},m_W) 
	  +6c_W^2B_0(q^2;m_W,m_W)-4c_W^2\Big\}  \\ \notag
	 &-\frac{g_Z^3}{2}m_Z\Big\{B_0(p^2_1;m_{h},m_{Z}) 
	 +B_0(p^2_2;m_{Z},m_{h}) \Big\} \\  \notag
	&+g^3m_W\Big\{
	 2c_W^2C_{hVV,1}^{VVV}(W,W,W)-2c_W^2C_{24}(m_{c^{\pm}},m_{c^{\pm}},m_{c^{\pm}})
	 +s_W^2C_{hVV,1}^{SVV}(G^{\pm},W,W) \\ \notag
	 &+s_W^2C_{hVV,1}^{VVS}(W,W,G^{\pm})
	 -2m_W^2\frac{s_W^4}{c_W^2}C_0(m_W,m_{G^{\pm}},m_W) \\\notag
	 &+\lambda_{hG^+G^-}v\frac{s_W^4}{c_W^2}C_0(m_{G^{\pm}},m_W,m_{G^{\pm}})  
	 -c_{2W}\frac{s_W^2}{c_W^2}\Big[C_{24}(m_W,m_{G^{\pm}},m_{G^{\pm}}) \\\notag
	 &+C_{24}(m_{G^{\pm}},m_{G^{\pm}},m_W) \Big] \Big\} \\ \notag
	 &+\frac{g_Z^3}{2}m_Z \Big\{
	 -2m_Z^2C_0(m_Z,m_h,m_Z)+6\lambda_{hhh}v C_{0}(m_h,m_Z,m_h) \\
	&+C_{24}(m_Z,m_h,m_{G^0})+C_{24}(m_{G^0},m_h,m_Z) \Big\},
\end{align}
\begin{align}
\notag
\label{3FB_hWW}
  16\pi^2 &\Gamma_{hWW}^{1,1\PI}(p^2_1,p^2_2,q^2)_B \\ \notag
  &=
  \frac{g^2}{4}\Big\{
2\lambda_{hH^{+}H^{-}}B_0(q^2;m_{H^{\pm}},m_{H^{\pm}})
+2\lambda_{hHH}B_0(q^2;m_H,m_H)+2\lambda_{hAA}B_0(q^2;m_A,m_A)
\notag\\\notag 
&-4\lambda_{hH^{+}H^{-}}C_{24}(m_{H^{\pm}},m_H,m_{H^{\pm}})
-4\lambda_{hH^{+}H^{-}}C_{24}(m_{H^{\pm}},m_A,m_{H^{\pm}})
\\ \nonumber
&-8\lambda_{hHH}C_{24}(m_H,m_{H^{\pm}},m_H)
-8\lambda_{hAA}C_{24}(m_A,m_{H^{\pm}},m_A)\Big\}\\\notag
\end{align}
\begin{align}
\notag
&+\frac{g^2}{4}
	\Big\{ 2\lambda_{hG^+G^-}B_0(q^2;m_{G^{\pm}},m_{G^{\pm}})
	 \\ \notag
	&+2\lambda_{hG^0G^0}B_0(q^2;m_{G^0},m_{G^0})
	 +6\lambda_{hhh}B_0(q^2;m_{h},m_{h}) \\ \notag
	 &-4\lambda_{hG^+G^-}C_{24}(m_{G^{\pm}},m_h,m_{G^{\pm}})
	 -4\lambda_{hG^+G^-}C_{24}(m_{G^{\pm}},m_{G^0},m_{G^{\pm}}) \\ \notag
	& -8\lambda_{hG^0G^0}C_{24}(m_{G^0},m_{G^{\pm}},m_{G^0})
	 -24\lambda_{hhh}C_{24}(m_{h},m_{G^{\pm}},m_{h}) \Big\}
	  \\ \notag
	  &-\frac{g^3}{2}m_W\Big\{6B_0(q^2;m_W,m_W)+6B_0(q^2;m_Z,m_Z)-8 \Big\} \\\notag 
	 &-\frac{g^3}{2}m_W\Big\{
	 B_0(p_1^2;m_h,m_W)+B_0(p^2_2;m_h,m_W)
	 +\frac{s_W^4}{c_W^2}B_0(p_1^2;m_{Z},m_{G^{\pm}}) \\\notag
	& +\frac{s_W^4}{c_W^2}B_0(p^2_2;m_Z,m_{G^{\pm}})  
	 +s_W^2B_0(p_1^2;m_{G^{\pm}},m_{\gamma})
	 +s_W^2B_0(p^2_2;m_{G^{\pm}},m_{\gamma}) \Big\} \\ \notag
	 &+g^3m_W\Big\{ C^{VVV}_{hVV,1}(Z,W,Z)
	 +c_W^2C^{VVV}_{hVV,1}(W,Z,W)
	 +s_W^2C^{VVV}_{hVV,1}(W,\gamma,W) \\ \notag
	 \ \ \ \ \ \ \ \ 
	 &-C_{24}(m_{c_Z},m_{c^{\pm}},m_{c_Z})-c_W^2C_{24}(m_{c^{\pm}},m_{c_Z},m_{c^{\pm}})
	 -s_W^2C_{24}(m_{c^{\pm}},m_{c_{\gamma}},m_{c^{\pm}})
	 \Big\} \\ \notag
	 &-\frac{g^3}{2}m_Ws_W^2\Big\{C_{hVV,1}^{SVV}(G^{\pm},Z,W)-C^{SVV}_ {hVV,1}(G^{\pm},\gamma,W)
	 +C_{hVV,1}^{VVS}(W,Z,G^{\pm}) \\\notag
	 &-C^{VVS}_ {hVV,1}(W,\gamma,G^{\pm})\Big\}  
	 -g^3m_W^3\Big\{C_0(m_W,m_h,m_W)+\frac{s_W^4}{c_W^4}C_0(m_Z,m_{G^{\pm}},m_Z) \Big\}	\\ \notag
	 &+\frac{g^3}{2}m_W\Big\{
	 \lambda_{hG^+G^-}v\frac{s_W^4}{c_W^2}C_0(m_{G^{\pm}},m_Z,m_{G^{\pm}})
	 + \lambda_{hG^+G^-}v s_W^2C_0(m_{G^{\pm}},m_{\gamma},m_{G^{\pm}}) \\\notag
	 &+6\lambda_{hhh}v C_0(m_h,m_W,m_h) 
	+ C_{24}(m_W,m_h,m_{G^{\pm}})+C_{24}(m_{G^{\pm}},m_h,m_W) \\
	&+\frac{s_W^2}{c_W^2}C_{24}(m_{G^0},m_{G^{\pm}},m_Z)
	+\frac{s_W^2}{c_W^2}C_{24}(m_Z,m_{G^{\pm}},m_{G^0})\Big\},
\end{align}
where
\begin{align}
\notag
C_{hVV,1}^{VVV}(X,Y,Z)&=[p^2_1(2C_{21}+3C_{11}+C_0)+p_2^2(2C_{22}+C_{12}) \\ 
&+p_1\cdot p_2(4C_{23}+3C_{12}+C_{11}-4C_0)+18C_{24}-3](m_X,m_Y,m_Z), \\ \notag
C_{hVV,1}^{SVV}(X,Y,Z)&=[p^2_1(C_{21}-C_0)+p_2^2(C_{22}-2C_{12}+C_0) \\ 
&+p_1\cdot p_2(2C_{23}-2C_{11})+3C_{24}-\frac{1}{2}](m_X,m_Y,m_Z), \\\notag
C_{hVV,1}^{VVS}(X,Y,Z)&=[p^2_1(C_{21}+4C_{11}+4C_0)+p_2^2(C_{22}+2C_{12}) \\ 
&+p_1\cdot p_2(2C_{23}+2C_{11}+4C_{12}+4C_0)+3C_{24}-\frac{1}{2}](m_X,m_Y,m_Z).
\end{align}

1PI diagram contributions to the scalar part of the $hff$ vertex is given by,
\begin{align}
\label{3F_hff_S}
\notag
\left(\frac{m_f}{v} \right)^{-1}16\pi^2& \Gamma^{S,1\PI}_{hff}(p^2_1,p^2_2,q^2)  \\ \notag
&=-2g_Z^4v^2(v_f^2-a_f^2)C_0(m_Z,m_f,m_Z)  
-\frac{g^2}{4}\{C_{SFV}(m_{G^{\pm}},m_{f^{\prime}},m_{W}) \\ \notag
&+C_{VFS}(m_W,m_{f^{\prime}},m_{G^{\pm}})\}  
-\frac{g_Z^2}{8}\{ C_{SFV}(m_{G^0},m_f,m_Z)+C_{VFS}(m_Z,m_f,m_{G^0})\} \\ \notag 
&+3\frac{m_h^2}{v}\frac{m_f^2}{v}C_0(m_h,m_f,m_h)
-\frac{m_h^2}{v}\frac{m_f^2}{v}C_0(m_{G^0},m_f,m_{G^0}) \\ \notag
&-2\frac{m_h^2}{v}\frac{m_{f^{\prime}}^2}{v}C_0(m_{G^{\pm}},m_{f^{\prime}},m_{G^{\pm}}) \\ \notag 
&-4g_Z^2(v^2_f-a_f^2)\{p^2_1(C_{21}+C_{11})+p^2_2(C_{22}+C_{12})+p_1\cdot p_2(2C_{23}-C_0) \\ \notag 
&+4C_{24}-1+m_f^2C_0\}(m_f,m_Z,m_f) \\ \notag 
&-4e^2Q_f^2\{p^2_1(C_{21}+C_{11})+p^2_2(C_{22}+C_{12})+p_1\cdot p_2(2C_{23}-C_0) \\ \notag 
&+4C_{24}-1+m_f^2C_0\}(m_f,m_{\gamma},m_f) \\ 
&-2\frac{m^2_{f^{\prime}}}{v^2}C^{FSF}_{hff}(f^{\prime},G^{\pm},f^{\prime})
-\frac{m^2_{f}}{v^2}C^{FSF}_{hff}(f,G^{0},f)
+\frac{m^2_{f}}{v^2}C^{FSF}_{hff}(f,h,f),
\end{align}
where
\begin{align}
\notag
C^{FSF}_{hff}(X,Y,Z)&=[m_X^2C_0+p^2_1(C_{11}+C_{21})+p^2_2(C_{12}+C_{22}) \\ 
&+2p_1\cdot p_2(C_{12}+C_{23})+4C_{24}](m_X,m_Y,m_Z)-\frac{1}{2},  \\ \notag
C^{VFS}_{hff}(X,Y,Z)&=[p^2_1(2C_0+3C_{11}+C_{21})+p^2_2(2C_{12}+C_{22}) \\ 
&+2p_1\cdot p_2(2C_0+2C_{11}+C_{12}+C_{23})+4C_{24}](m_X,m_Y,m_Z)-\frac{1}{2}, \\ \notag
C^{SFV}_{hff}(X,Y,Z)&=[p^2_1(C_{21}-C_0)+p^2_2(C_{22}-C_{12}) \\ 
&+2p_1\cdot p_2(C_{23}-C_{12})+4C_{24}](m_X,m_Y,m_Z)-\frac{1}{2}.
\end{align}

Finally, we attach the formulae for the decay rate of $h\to \gamma\gamma$ in the IDM; 
\begin{equation}
\Gamma (h\to\gamma \gamma) = \frac{\sqrt{2}G_F\alpha_{em}^2m_h^3}{64\pi^3}\left| -I_{H^{\pm}}[m_h^2]+\sum_fQ_f^2N_c^fI_f[m_h^2]+I_W[m_h^2] \right|^2,
\end{equation}
where the parts of the inert scalar loop, the fermion loop and the gauge boson loop are given by 
\begin{align}
\label{hgamgam_s}
I_{H^{\pm}}[m_h^2] &= \frac{\upsilon \lambda_{hH^{+}H^{-}}}{m_h^2}[1+2m_{H^{\pm}}^2C_0(0,0,m_h^2;m_{H^{\pm}},m_{H^{\pm}},m_{H^{\pm}})], \\ 
\label{hgamgam_f}
I_f[m_h^2]&=-\frac{4m_f^2}{m_h^2}\left[1-\frac{m_h^2}{2}\left(1-\frac{4m_f^2}{m_h^2}\right)C_0(0,0,m_h^2;m_f,m_f,m_f)\right], \\
\label{hgamgam_V}
I_W[m_h^2]&=1+\frac{6m_W^2}{m_h^2}-6m_W^2\left(1-\frac{2m_W^2}{m_h^2}\right)C_0(0,0,m_h^2;m_f,m_f,m_f). 
\end{align}

\color{black}

\end{document}